\documentclass[12pt,thmsa]{article}%
\usepackage{amssymb}
\usepackage{sw20lart}
\usepackage[singlespacing]{setspace}
\usepackage{amsmath}
\usepackage{amsfonts}
\usepackage{graphicx}
\usepackage{cite}%
\allowdisplaybreaks
\begin{document}

\author{\vspace{0.16in}Hartmut Wachter\thanks{e-mail:
Hartmut.Wachter@physik.uni-muenchen.de}\\Max-Planck-Institute\\for Mathematics in the Sciences \\Inselstr. 22, D-04103 Leipzig, Germany\\\hspace{0.4in}\\Arnold-Sommerfeld-Center\\Ludwig-Maximilians-Universit\"{a}t\\Theresienstr. 37, D-80333 M\"{u}nchen, Germany}
\title{Braided products for quantum spaces}
\date{}
\maketitle

\begin{abstract}
Attention is focused on quantum spaces of physical importance, i.e. Manin
plane, q-deformed Euclidean space in three or four dimensions as well as
q-deformed Minkowski space. There are algebra isomorphisms that allow to
identify quantum spaces with commutative coordinate algebras. This observation
enables the realization of braiding mappings on commutative algebras. The aim
of the article is to perform this task explicitly, leading us to formulae for
so-called braided products. \newpage

\end{abstract}

\section{Introduction}

Relativistic quantum field theory is not a fundamental theory, since its
formalism leads to divergencies. In some cases like that of quantum
electrodynamics one is able to overcome the difficulties with the divergencies
by applying the so-called renormalization procedure due to Richard Feynman.
Unfortunately, this procedure is not successful if we want to deal with
quantum gravity. Despite the fact that gravitation is a rather weak
interaction we are not able to treat it perturbatively. The reason for this
lies in the fact, that transition amplitudes of nth order to the gravitation
constant diverge like a momentum integral of the general form \cite{Wei}%
\begin{equation}
\int p^{2n-1}dp,\quad n\in\mathbb{N}. \label{ImInt}%
\end{equation}
This leaves us with an infinite number of ultraviolet divergent Feynman
diagrams that cannot be removed by redefining finitely many physical parameters.

It is surely legitimate to ask for the reason for these fundamental
difficulties. It is commonplace that the problems with the divergences in
relativistic quantum field theory result from an incomplete description of
spacetime at very small distances. Niels Bohr and Werner Heisenberg have been
the first who suggested that quantum field theories should be formulated on a
spacetime lattice \cite{Cass, Heis38}. Such a spacetime lattice would imply
the existence of a smallest distance $a$ with the consequence that plane-waves
of wave-length smaller than twice the lattice spacing could not propagate. In
accordance with the relationship between wave-length $\lambda$ and momentum
$p$ of a plane-wave, i.e.
\begin{equation}
\lambda\geq\lambda_{\min}=2a\quad\Rightarrow\quad\frac{1}{\lambda}\sim p\leq
p_{\max}\sim\frac{1}{2a},
\end{equation}
it then follows that physical momentum space would be bounded. Hence, the
domain of all momentum integrals in Eq. (\ref{ImInt}) would be bounded as
well. Consequently, all of these integrals should take on finite values.

However, discrete spacetime structures in general do not respect classical
Poincar\'{e} symmetry. A possible way out of this difficulty is to modify not
only spacetime but also its corresponding symmetries. How are we to accomplish
this? First of all let us recall that classical spacetime symmetries are
usually described by Lie groups. Realizing that Lie groups are manifolds the
Gelfand-Naimark\textsf{\ }theorem tells us that Lie groups can be naturally
embedded in the category of algebras\textsf{\ }\cite{GeNe}. The utility of
this interrelation lies in formulating the geometrical structure of Lie groups
in terms of a Hopf structure \cite{Hopf}. The point is that during the last
two decades generic methods have been discovered for continuously deforming
matrix groups and Lie algebras within the category of Hopf algebras. It is
this development which finally led to the arrival of quantum groups and
quantum spaces \cite{Ku83, Wor87, Dri85, Jim85, Drin86, RFT90, Tak90}.

From a physical point of view the most realistic and interesting deformations
are given by q-deformed versions of Minkowski space and Euclidean spaces
together with their corresponding symmetries, i.e. respectively Lorentz
symmetry and rotational symmetry \cite{CSSW90, PW90, SWZ91, Maj91, LWW97}.
Further studies even allowed to establish differential calculi on these
q-deformed quantum spaces \cite{WZ91, CSW91, Song92} representing nothing
other than q-analogs of classical translational symmetry. In this sense we can
say that q-deformations of the complete Euclidean and Poincar\'{e} symmetries
are now available \cite{OSWZ92, Maj93-2}. Finally, Julius Wess and his
coworkers were able to show that q-deformation of spaces and symmetries can
indeed lead to the wanted discretizations of the spectra of spacetime
observables \cite{FLW96, Wes97, CW98}. This observation nourishes the hope
that q-deformation might give a new method to regularize quantum field
theories \cite{GKP96, MajReg, Oec99, Blo03}.

In order to formulate quantum field theories on q-deformed quantum spaces it
is necessary to provide us with some essential tools of a q-deformed analysis.
The main question is how to define these new tools, which should be q-analogs
of classical notions. Towards this end the considerations of Shahn Majid have
proven very useful \cite{Maj91Kat, Maj94Kat, Maj93-6}. The key idea of his
approach is that all the quantum spaces to a given quantum symmetry form a
braided tensor category. Consequently, operations and objects concerning
quantum spaces must rely on this framework of a braided tensor category, in
order to guarantee their well-defined behavior under quantum group
transformations. This so-called principle of covariance can be seen as the
essential guideline for constructing a consistent theory.

In our previous work we have applied these rather general considerations (as
they are exposed in Refs. \cite{Maj93-6, Maj94-10, Maj93-5, Maj95}) to quantum
spaces of physical importance, i.e. q-deformed quantum plane, q-deformed
Euclidean space with three or four dimensions and q-deformed Minkowski space,
resulting in explicit expressions for star products \cite{WW01}, operator
representations \cite{BW01}, q-integrals \cite{Wac02}, q-exponentials
\cite{Wac03}, and q-translations \cite{Wac04}. In this article we would like
to complete our toolbox of essential elements of q-analysis by calculating
formulae for braided products.

For this to achieve we intend to proceed as follows. In Sec. \ref{BasSec} we
cover the ideas our considerations about q-analysis are based on. Important
for us is the fact that there is an algebra isomorphism which allows to
identify\ quantum space algebras with algebras of commutative coordinates. In
the subsequent sections this observation will be used to calculate formulae
for braided products, i.e. realizations of braiding mappings on commutative
algebras. In doing so, we restrict attention on braidings referring to quantum
spaces we are interested in for physical reasons, i.e. Manin plane, q-deformed
Euclidean space in three or four dimensions as well as q-deformed Minkowski
space. In Sec. \ref{Conclusion} we close our considerations by a short
conclusion. Appendix \ref{QuanAlg} shall serve as a review of the quantum
algebras describing the symmetries of the quantum spaces under consideration,
while Appendix \ref{ActSymAlg} is devoted to some special calculations.

\section{Basic ideas on q-analysis\label{BasSec}}

Let us recall that q-analysis can be regarded as a noncommutative analysis
formulated within the framework of quantum spaces \cite{Wess00}. Quantum
spaces are defined as comodule algebras of quantum groups and\ can be
interpreted as deformations of ordinary coordinate algebras. For our purposes,
it is at first sufficient to consider a quantum space as an algebra
$\mathcal{A}_{q}$ of formal power series in non-commuting coordinates
$X_{1},X_{2},\ldots,X_{n},$ i.e.%

\begin{equation}
\mathcal{A}_{q}=\mathbb{C}\left[  \left[  X_{1},\ldots X_{n}\right]  \right]
/\mathcal{I},
\end{equation}
where $\mathcal{I}$ denotes the ideal generated by the relations of the
non-commuting coordinates. The three-dimensional Euclidean quantum space shall
serve as an example. It consists of all the power series in three coordinates
$X^{+},$ $X^{3},$ and $X^{-}$ being subject to the commutation relations
\begin{align}
X^{3}X^{\pm}-q^{\pm2}X^{\pm}X^{3}  &  =0,\label{3dim KoorRel}\\
X^{-}X^{+}-X^{+}X^{-}  &  =\lambda X^{3}X^{3}.\nonumber
\end{align}
We can think of $q>1$ as a deformation parameter measuring the coupling among
different spatial degrees of freedom, whereas $\lambda\equiv q-q^{-1}$ should
here be understood as some kind of a relative lattice spacing. In the
classical case, i.e. if $q$ tends to $1,$ it is rather obvious from Eq.
(\ref{3dim KoorRel}) that we regain commutative coordinates.

Next, we would like to focus our attention on the question how to perform
calculations on our noncommutative coordinate algebras. As will become
evident, this can be accomplished by a kind of pullback, which translates
operations on noncommutative coordinate algebras to those on commutative ones.
Towards this end, we have to realize that the noncommutative algebra
$\mathcal{A}_{q}$ satisfies the \textit{Poincar\'{e}-Birkhoff-Witt property},
i.e. the dimension of a subspace of homogenous polynomials should be the same
as for commuting coordinates. This property is the deeper reason why the
monomials of normal ordering $X_{1}X_{2}\ldots X_{n}$ constitute a basis of
$\mathcal{A}_{q}$. In particular, we can establish a vector space isomorphism
between $\mathcal{A}_{q}$ and the commutative algebra $\mathcal{A}$ generated
by ordinary coordinates $x_{1},x_{2},\ldots,x_{n}$:
\begin{align}
\mathcal{W}  &  :\mathcal{A}\longrightarrow\mathcal{A}_{q},\label{AlgIso}\\
\mathcal{W}(x_{1}^{i_{1}}\ldots x_{n}^{i_{n}})  &  =X_{1}^{i_{1}}\ldots
X_{n}^{i_{n}}.\nonumber
\end{align}

This vector space isomorphism can even be extended to an algebra isomorphism
by introducing a noncommutative product in $\mathcal{A}$, the so-called
\textit{star product} \cite{Moy49, BFF78, MSSW00}. This product is defined via
the relation
\begin{equation}
\mathcal{W}(f\star g)=\mathcal{W}(f)\,\mathcal{W}(g), \label{StarDef}%
\end{equation}
being tantamount to%
\begin{equation}
f\star g=\mathcal{W}^{-1}\left(  \mathcal{W}\left(  f\right)  \,\mathcal{W}%
\left(  g\right)  \right)  ,
\end{equation}
where $f$ and $g$ are formal power series in $\mathcal{A}$. In the case of the
three-dimensional q-deformed Euclidean space, for instance, the star product
takes the form%
\begin{align}
&  f(\underline{x})\star g(\underline{x})\label{sternformel}\\
&  =\sum_{i=0}^{\infty}\lambda^{i}\frac{(x^{3})^{2i}}{[[i]]_{q^{4}}%
!}\,q^{2(\hat{n}_{3}\hat{n}_{+}^{\prime}+\,\hat{n}_{-}\hat{n}_{3}^{\prime}%
)}\left.  \left[  (D_{q^{4}}^{-})^{i}f(\underline{x})\right]  \left[
(D_{q^{4}}^{+})^{i}g(\underline{x}^{\prime})\right]  \right\vert _{x^{\prime
}\rightarrow x}\nonumber\\
&  =f(\underline{x})\,g(\underline{x})+O(h),\nonumber
\end{align}
where
\begin{equation}
h\equiv\ln q\text{\quad and\quad}\hat{n}_{A}\equiv\frac{\partial}{\partial
x^{A}},\quad A\in\{+,3,-\}.
\end{equation}

In addition to this, we have introduced in (\ref{sternformel}) as some sort of
discretised version of classical derivatives the so-called \textit{Jackson
derivative} \cite{Jack08}%
\begin{equation}
D_{q^{a}}^{A}f(x^{A})=\frac{f(q^{a}x^{A})-f(x^{A})}{q^{a}x^{A}-x^{A}},\quad
a\in\mathbb{C}.
\end{equation}
Furthermore, the \textit{antisymmetric q-numbers} are given by%
\begin{equation}
\left[  \left[  n\right]  \right]  _{q^{a}}\equiv\sum_{k=0}^{n-1}q^{ak}%
=\frac{1-q^{an}}{1-q^{a}},
\end{equation}
showing the property%
\begin{equation}
\left[  \left[  n\right]  \right]  _{q^{a}}\rightarrow n\quad\text{for
}q\rightarrow1.
\end{equation}
Factorials of q-numbers are defined in complete analogy to the classical case,
i.e.%
\begin{equation}
\left[  \left[  m\right]  \right]  _{q^{a}}!\equiv\left[  \left[  1\right]
\right]  _{q^{a}}\left[  \left[  2\right]  \right]  _{q^{a}}\ldots\left[
\left[  m\right]  \right]  _{q^{a}},\qquad\left[  \left[  0\right]  \right]
_{q^{a}}!\equiv1.
\end{equation}
From these expressions it should be evident that star products on q-deformed
quantum spaces tend to the commutative product in the limit $q\rightarrow1,$
i.e. our star products can be seen as modifications of commutative products.

Now, we want to deal on with symmetry generators. To this end let us recall
that in quantum theory it is necessary to know how symmetry generators act on
function spaces. Such actions can be derived from the commutation relations
between symmetry generators and space coordinates. As remarked earlier, there
are q-deformed analogs of classical symmetry algebras and we also know the
commutation relations of the corresponding generators with quantum space
coordinates. In the case of three-dimensional q-deformed Euclidean space, for
example, quantum space coordinates now commute with angular momentum component
$L^{+}$ according to
\begin{align}
L^{+}X^{+}-X^{+}L^{+}  &  =0,\label{VerLX}\\
L^{+}X^{3}-X^{3}L^{+}  &  =qX^{+}\tau^{-1/2},\nonumber\\
L^{+}X^{-}-X^{-}L^{+}  &  =X^{3}\tau^{-1/2},\nonumber
\end{align}
and the partial derivative $\partial_{-}$ has to obey the q-deformed
Heisenberg relations
\begin{align}
\partial_{-}X^{+}-X_{-}\partial^{+}  &  =0,\label{explizit1}\\
\partial_{-}X^{3}-q^{2}X^{3}\partial_{-}  &  =-q^{2}\lambda\lambda_{+}%
X^{+}\partial_{3},\nonumber\\
\partial_{-}X^{-}-q^{4}X^{-}\partial_{-}  &  =q^{2}\lambda\lambda_{+}%
X^{3}\partial_{3}+q\lambda^{2}\lambda_{+}X^{+}\partial_{+},\nonumber
\end{align}
with $\lambda_{+}\equiv q+q^{-1}.$

From such identities one can calculate actions of q-deformed symmetry
generators on normal ordered monomials of quantum space coordinates. By means
of the relation
\begin{equation}
\mathcal{W}(h\triangleright f)=h\triangleright\mathcal{W}(f),\quad
h\in\mathcal{H}\text{, }f\in\mathcal{A}\text{,}%
\end{equation}
or%
\begin{equation}
h\triangleright f=\mathcal{W}^{-1}\left(  h\triangleright\mathcal{W}%
(f)\right)  ,
\end{equation}
the action of the symmetry algebra $\mathcal{H}$ on the quantum space algebra
$\mathcal{A}_{q}$ carries over to a corresponding one on a commutative
coordinate algebra $\mathcal{A}$. We applied these ideas in the work of Ref.
\cite{BW01}. This way we got operator representations of q-deformed symmetry
generators. As an example we give the representations that result from the
relations in (\ref{VerLX}) and (\ref{explizit1}):%
\begin{align}
L^{+}\triangleright f(\underline{x})  &  =-q^{2}x^{3}(D_{q^{4}}^{-}%
f)(q^{-2}x^{-})-qx^{+}(D_{q^{2}}^{3}f)(q^{-2}x^{-}),\label{AblD-}\\
\partial_{-}\triangleright f(\underline{x})  &  =D_{q^{4}}^{-}f(q^{2}%
x^{3})+\lambda x^{+}(D_{q^{2}}^{3})^{2}f.\nonumber
\end{align}
In this manner, we obtained discretised versions of classical expressions.
Furthermore, let us note that the non-classical terms proportional to
$\lambda$ can again be interpreted as result of the coupling between the
different degrees of freedom. So to speak, the situation in quantum spaces is
in some sense like that in solids.

As a next step we can introduce q-deformed integrals as inverse operations to
partial derivatives \cite{Wac02}. This requires to find solutions of the
difference equations
\begin{equation}
\partial_{A}\triangleright F=f
\end{equation}
for given $f.$ As an example we present the integral corresponding to the
second identity in (\ref{AblD-}):%
\begin{align}
F  &  =(\partial_{-})^{-1}\triangleright f\\
&  =\sum_{k=0}^{\infty}(-\lambda)^{k}q^{2k(k+1)}\left[  x^{+}(D_{q^{4}}%
^{-})^{-1}(D_{q^{2}}^{3})^{2}\right]  ^{k}(D_{q^{4}}^{-})^{-1}f(q^{-2(k+1)}%
x^{3}).\nonumber
\end{align}
The so-called \textit{Jackson integral }in here is defined by%
\begin{equation}
\left.  (D_{q^{a}}^{A})^{-1}f\right\vert _{0}^{x^{A}}=-(1-q^{a})\sum
_{k=1}^{\infty}(q^{-ak}x^{A})f(q^{-ak}x^{A}).
\end{equation}
Again, it can be seen as discretised version of classical integrals, into
which it passes for $q\rightarrow1$.

Now, we come to another very important ingredient of q-analysis. As we know
from quantum field theory, it is convenient to express solutions to wave
equations in terms of plane-waves, i.e. exponential functions. In q-analysis
there are q-deformed analogs of such exponentials, which are completely
determined by their property of being eigenfunctions of partial derivatives,
i.e.
\begin{equation}
\exp(x\mid\partial)\in\mathcal{A}_{x}\otimes\mathcal{A}_{\partial},
\end{equation}
with
\begin{equation}
\partial_{A}\triangleright\exp(x\mid\partial)=\exp(x\mid\partial)\star
\partial_{A}\quad\text{for all }A.
\end{equation}
In some cases like that of three-dimensional q-deformed Euclidean space these
exponentials take on a form that is in complete analogy to their classical
counterpart \cite{Wac03}:%
\begin{align}
&  \exp(x_{\bar{R}}\mid\partial_{L})=\\
&  \sum_{\underline{n}=0}^{\infty}\frac{(x^{+})^{n_{+}}(x^{3})^{n_{3}}%
(x^{-})^{n_{-}}\otimes(\partial_{-})^{n_{-}}(\partial_{3})^{n_{3}}%
(\partial_{+})^{n_{+}}}{[[n_{+}]]_{q^{4}}![[n_{3}]]_{q^{2}}![[n_{-}]]_{q^{4}%
}!}.\nonumber
\end{align}

The q-deformed exponentials can be used to perform translations on quantum
spaces, as it was explained in Ref. \cite{Maj93-5}. In complete analogy to the
classical case, translations on quantum spaces are obtained by the formula
\begin{equation}
f(y\,\underline{\oplus}_{\bar{L}}x)\equiv\exp(y_{\bar{R}}\mid\partial
_{L})\triangleright f(\underline{x}),
\end{equation}
with the understanding that the partial derivatives act on the function far
right in the above equation. By using the last identity in combination with
the explicit form for q-exponential and actions of partial derivatives we can
obtain q-deformed versions of Taylor rules. As an example we give the result
for three-dimensional q-deformed Euclidean space \cite{Wac04}:%
\begin{align}
&  f(y\,\underline{\oplus}_{\bar{L}}\,x)\\
&  =\sum_{k_{+}=0}^{n_{+}}\sum_{k_{3}=0}^{n_{3}}\sum_{k_{-}=0}^{n_{-}}%
\sum_{i=0}^{k_{3}}\frac{(q\lambda\lambda_{+})^{i}}{[[2i]]_{q^{2}}!!}%
\frac{(y^{+})^{k_{+}}(y^{3})^{k_{3}-i}(y^{-})^{k_{-}+\,i}}{[[k_{+}]]_{q^{4}%
}![[k_{3}-i]]_{q^{2}}![[k_{-}]]_{q^{4}}!}\nonumber\\
&  \times\,(x^{+})^{i}\Big ((D_{q^{4}}^{+})^{k_{+}}(D_{q^{2}}^{3})^{k_{3}%
+i}(D_{q^{4}}^{-})^{k_{-}}f\Big )(q^{2(k_{3}-i)}x^{+},q^{2k_{-}}%
x^{3}).\nonumber
\end{align}

Now, we come to the main incentive of this paper, i.e. braided products. To
this end, let us recall that in physical applications it is often necessary to
deal with tensor products. Such a situation arises, for example, when one
wants to describe multiparticle states or the phenomenon of spin. The tensor
product of commutative algebras is usually characterized by a componentwise
multiplication. For tensor products of q-deformed quantum spaces, however,
this is not the case.

For this to become more clear, it is useful to consider quantum spaces from a
point of view provided by category theory. A category for our purposes is just
a collection of objects $X,Y,Z,\ldots$ and a set Mor$(X,Y)$ of morphisms
between any two objects $X,Y.$ The composition of morphisms has similar
properties as the composition of maps. We are interested in tensor categories.
These categories have a product, denoted $\otimes$ and called the tensor
product. It admits several 'natural' properties such as associativity and
existence of a unit object. For a more formal treatment we refer to Refs.
\cite{Maj91Kat}, \cite{Maj94Kat}, \cite{Maj95} or \cite{MaL74}. If the action
of a quasitriangular Hopf algebra $\mathcal{H}$ on the tensor product of two
quantum spaces $X$ and $Y$ is defined by
\begin{equation}
h\triangleright(v\otimes w)=(h_{(1)}\triangleright v)\otimes(h_{(2)}%
\triangleright w)\in X\otimes Y,\quad h\in\mathcal{H}, \label{WirkHopfTens}%
\end{equation}
(the coproduct is written in the so-called Sweedler notation, i.e.
$\Delta(h)=h_{(1)}\otimes h_{(2)}$) then the representations (quantum spaces)
of the given Hopf algebra (quantum algebra) are the objects of a tensor category.

In this tensor category exists a morphism of particular importance. To be more
precise, for any pair of objects $X,Y$ there is an isomorphism $\Psi
_{X,Y}:X\otimes Y\rightarrow Y\otimes X$ such that $(g\otimes f)\circ
\Psi_{X,Y}=\Psi_{X^{\prime},Y^{\prime}}\circ(f\otimes g)$ for arbitrary
morphisms $f\in$ Mor$(X,X^{\prime})$ and $g\in$ Mor$(Y,Y^{\prime})$. In
addition to this one can require the hexagon axiom to hold. The hexagon axiom
is the validity of the two conditions
\begin{equation}
\Psi_{X,Z}\circ\Psi_{Y,Z}=\Psi_{X\otimes Y,Z},\quad\Psi_{X,Z}\circ\Psi
_{X,Y}=\Psi_{X,Y\otimes Z}.
\end{equation}
A tensor category equipped with such mappings $\Psi_{X,Y}$ for each pair of
objects $X,Y$ is called a braided tensor category. The mappings $\Psi_{X,Y}$
as a whole are often referred to as the braiding of the tensor category. Their
property of being morphisms implies that it makes no difference whether we
interchange the two tensor factors by the braiding before or after applying
the action of a given Hopf algebra. More formally, we have%
\begin{equation}
\Psi_{X,Y}\left(  h\triangleright(v\otimes w)\right)  =h\triangleright\left(
\Psi_{X,Y}(v\otimes w)\right)  \in Y\otimes X,\quad h\in\mathcal{H}.
\end{equation}
To state this another way, the braiding represents a solution to the problem
how to commute elements of different quantum spaces in a covariant way, i.e.
consistent with the underlying quantum symmetry.

One should notice that the above considerations also hold for the set of
inverse mappings $\Psi_{X,Y}^{-1}$. This fact gives rise to a second braiding,
which in our cases is indeed different from the first one. In the sequel of
this article, it is our aim to derive explicit formulae for the mappings
$\Psi_{X,Y}$ and $\Psi_{X,Y}^{-1}$, where $X$ and $Y$ stand for quantum space
algebras of physical importance, i.e. Manin plane, q-deformed Euclidean space
in three or four dimensions as well as q-deformed Minkowski space. In doing
so, it can be useful to realize that the symmetry algebras for these quantum
spaces are described by quasitriangular Hopf algebras. Thus, the braiding in
the corresponding tensor category is completely determined by a universal
R-matrix $\mathcal{R=R}^{(1)}\otimes\mathcal{R}^{(2)}\in\mathcal{H}%
\otimes\mathcal{H}$ (or its inverse $\mathcal{R}^{-1}\mathcal{=}%
(\mathcal{R}^{-1}\mathcal{)}^{(1)}\otimes(\mathcal{R}^{-1}\mathcal{)}^{(2)}$),
since we have%
\begin{align}
\Psi_{X,Y}(v\otimes w)  &  =(\mathcal{R}^{(2)}\triangleright w)\otimes
(\mathcal{R}^{(1)}\triangleright v),\quad v\in X,\;w\in Y,\label{AllBraiRel}\\
\Psi_{X,Y}^{-1}(v\otimes w)  &  =((\mathcal{R}^{-1}\mathcal{)}^{(2)}%
\triangleright w)\otimes((\mathcal{R}^{-1}\mathcal{)}^{(1)}\triangleright
v).\nonumber
\end{align}
By means of the algebra isomorphism in Eq. (\ref{AlgIso}) we are able to
introduce a new product between two commutative algebras $\mathcal{A}$ and
$\mathcal{A}^{\prime}$, the so-called \textit{braided product}, which for the
two different braidings is given by%
\begin{align}
\underline{\odot}_{L},\underline{\odot}_{\bar{L}}  &  :\mathcal{A}%
\otimes\mathcal{A}^{\prime}\rightarrow\mathcal{A}^{\prime}\otimes
\mathcal{A},\label{AllZopPr}\\[0.16in]
f(x)\,\underline{\odot}_{L}\,g(y)  &  \equiv\left(  \mathcal{W}^{-1}%
\otimes\mathcal{W}^{-1}\right)  \circ\Psi_{X,Y}^{-1}\left(  \mathcal{W}%
(f)\otimes\mathcal{W}(g)\right)  ,\nonumber\\
f(x)\,\underline{\odot}_{\bar{L}}\,g(y)  &  \equiv\left(  \mathcal{W}%
^{-1}\otimes\mathcal{W}^{-1}\right)  \circ\Psi_{X,Y}\left(  \mathcal{W}%
(f)\otimes\mathcal{W}(g)\right)  .\nonumber
\end{align}

Finally, it should be mentioned that braided products are part of the
calculation of q-deformed tensor products. For this to become more clear, let
us recall the definition for a tensor product of two quantum spaces. In the
case of the braiding $\Psi_{X,Y}$ it reads%
\begin{gather}
(X\otimes Y)\otimes(X\otimes Y)\rightarrow X\otimes Y\label{DefBraiTens}\\
(w_{1}\otimes v_{1})(w_{2}\otimes v_{2})\equiv(m\otimes m)\circ(w_{1}%
\otimes\Psi_{X,Y}(v_{1}\otimes w_{2})\otimes v_{2}),\nonumber
\end{gather}
and likewise for the inverse braiding $\Psi_{X,Y}^{-1}$, where $m$ denotes
multiplication in the quantum space algebras $X$ and $Y$. With this identity
at hand, it is now straightforward to extend the definition of braided
products to that of a q-deformed\ tensor product between the algebras
$\mathcal{A}$ and $\mathcal{A}^{\prime}$:%
\begin{align}
&  \underline{\odot}_{\bar{L}}:(\mathcal{A}\otimes\mathcal{A}^{\prime}%
)\otimes(\mathcal{A}\otimes\mathcal{A}^{\prime})\rightarrow\mathcal{A}%
\otimes\mathcal{A}^{\prime}\label{BraidTens}\\[0.16in]
&  \left(  f_{1}(x)\otimes g_{1}(y)\right)  {}\,\underline{\odot}_{\bar{L}%
}\,{}\left(  f_{2}(x)\otimes g_{2}(y)\right) \nonumber\\
&  \equiv\left(  \mathcal{W}^{-1}\otimes\mathcal{W}^{-1}\right)
\circ(m\otimes m)\circ\big[\mathcal{W}(f_{1})\otimes\Psi\left(  \mathcal{W}%
(g_{1})\otimes\mathcal{W}(f_{2})\right)  \otimes\mathcal{W(}g_{2}%
)\big]\nonumber\\
&  =\left[  f_{1}\star\mathcal{W}^{-1}\left(  \mathcal{R}^{(2)}\triangleright
\mathcal{W}(f_{2})\right)  \right]  \otimes\left[  \mathcal{W}^{-1}\left(
\mathcal{R}^{(1)}\triangleright\mathcal{W}(g_{1})\right)  \star g_{2}\right]
\nonumber\\
&  =f_{1}(x)\star_{x}\left(  g_{1}(y)\,\underline{\odot}_{\bar{L}}%
\,f_{2}(x)\right)  \star_{y}g_{2}(y).\nonumber
\end{align}
It is not very hard to adapt this formula\ to the inverse braiding. Thus, the
details are left to the reader.

\section{Quantum plane in two dimensions}

In this section we would like to consider the simplest example for a
q-deformed quantum space, the so-called Manin plane \cite{Man88}. It is
generated by coordinates $X^{\alpha},$ $\alpha=1,2,$ subjected to the relation%
\begin{equation}
X^{1}X^{2}=qX^{2}X^{1}.
\end{equation}
The corresponding symmetry algebra is given by $U_{q}(su_{2})$ \cite{KR81}
(for its definition see also Appendix \ref{QuanAlg}). If we substitute for the
universal R-matrix that of $U_{q}(su_{2})$ and take into account the action of
$U_{q}(su_{2})$ on the quantum plane, the relations in Eq. (\ref{AllBraiRel})
enable us to calculate the braiding between two copies of quantum plane
coordinates. Proceeding in this way one can obtain \cite{Mik04}%
\begin{align}
\Psi_{X,Y}\left(  X^{\alpha}\otimes Y^{\beta}\right)   &  =qR\,_{\gamma\delta
}^{\alpha\beta}\,Y^{\gamma}\otimes X^{\delta},\\
\Psi_{X,Y}^{-1}\left(  X^{\alpha}\otimes Y^{\beta}\right)   &  =q^{-1}%
(R^{-1})_{\gamma\delta}^{\alpha\beta}\,Y^{\gamma}\otimes X^{\delta},\nonumber
\end{align}
where $R\,_{\gamma\delta}^{\alpha\beta}$ and $(R^{-1})_{\gamma\delta}%
^{\alpha\beta}$ stand for the vector representation of the universal R-matrix
and its inverse, respectively. More explicitly, we have the relations%
\begin{align}
\Psi_{X,Y}\left(  X^{1}\otimes Y^{1}\right)   &  =q^{2}\,Y^{1}\otimes
X^{1},\label{BraidKoord2dim}\\
\Psi_{X,Y}\left(  X^{1}\otimes Y^{2}\right)   &  =q\,Y^{2}\otimes
X^{1}+q\lambda Y^{1}\otimes X^{2},\nonumber\\
\Psi_{X,Y}\left(  X^{2}\otimes Y^{1}\right)   &  =q\,Y^{1}\otimes
X^{2},\nonumber\\
\Psi_{X,Y}\left(  X^{2}\otimes Y^{2}\right)   &  =q^{2}\,Y^{2}\otimes
X^{2}.\nonumber
\end{align}
The relations arising from the inverse braiding are derived from the above
ones most easily by applying the substitutions%
\begin{equation}
X^{\alpha}\leftrightarrow X^{\alpha^{\prime}},\quad Y^{\alpha}\leftrightarrow
Y^{\alpha^{\prime}},\quad q\leftrightarrow q^{-1},\quad\alpha^{\prime}%
\equiv3-\alpha,\quad\alpha=1,2. \label{TranRul2dim}%
\end{equation}

Now, we come to the calculation of braided products for\ two normal ordered
monomials. Towards this end, it suffices to focus our attention on one version
of the braiding, since the expressions for the two braidings follow from each
other quite easily by using simple transition rules [these transition rules
are based on the correspondence in Eq. (\ref{TranRul2dim}), as we will see
later on].

Using Eq. (\ref{BraidKoord2dim}) together with Eq. (\ref{TranRul2dim}), it is
a simple exercise to check that%
\begin{align}
\Psi_{X,Y}^{-1}\left(  (X^{1})^{n_{1}}\otimes(Y^{1})^{m_{1}}\right)   &
=q^{-2n_{1}m_{1}}(Y^{1})^{m_{1}}\otimes(X^{1})^{n_{1}}, \label{Psi-12dimKoord}%
\\
\Psi_{X,Y}^{-1}\left(  (X^{2})^{n_{2}}\otimes(Y^{2})^{m_{2}}\right)   &
=q^{-2n_{2}m_{2}}(Y^{2})^{m_{2}}\otimes(X^{2})^{n_{2}},\nonumber\\
\Psi_{X,Y}^{-1}\left(  (X^{1})^{n_{1}}\otimes(Y^{2})^{m_{2}}\right)   &
=q^{-n_{1}m_{2}}(Y^{2})^{m_{2}}\otimes(X^{1})^{n_{1}}.\nonumber
\end{align}
To derive a formula for the braiding between powers of $X^{2}$ and $Y^{1}$ is
a little bit more complicated. A glance at the second relation in Eq.
(\ref{BraidKoord2dim}) shows that we can make as ansatz%
\begin{align}
&  \Psi_{X,Y}^{-1}\left(  (X^{2})^{n_{2}}\otimes(Y^{1})^{m_{1}}\right)
\label{RekBrai}\\
&  =\sum_{i=0}^{\min(n_{2},m_{1})}C\,_{i}^{n_{2},m_{1}}\,(Y^{1})^{m_{1}%
-i}(Y^{2})^{i}\otimes(X^{1})^{i}(X^{2})^{n_{2}-i},\nonumber
\end{align}
with unknown coefficients $C\,_{i}^{n,m}\in\mathbb{C}.$ Exploiting Eq.
(\ref{BraidKoord2dim}) together with Eq. (\ref{TranRul2dim}) should enable us
to find the recursion relation%
\begin{align}
C\,_{i}^{n,m}  &  =q^{-n-2i}C\,_{i}^{n,m-1}-q^{-n}\lambda\,[[n-i+1]]_{q^{-2}%
}\,C\,_{i-1}^{n,m-1},\\
C\,_{0}^{n,m}  &  =q^{-nm}.\nonumber
\end{align}
As one can prove by insertion this recursion relation has the solution%
\begin{equation}
C\,_{i}^{n,m}=q^{-nm}(-\lambda)^{i}[[i]]_{q^{-2}}!%
%TCIMACRO{\QATOPD{[}{]}{n}{i}}%
%BeginExpansion
\genfrac{[}{]}{0pt}{}{n}{i}%
%EndExpansion
_{q^{-2}}%
%TCIMACRO{\QATOPD{[}{]}{m}{i}}%
%BeginExpansion
\genfrac{[}{]}{0pt}{}{m}{i}%
%EndExpansion
_{q^{-2}},
\end{equation}
where the \textit{q-deformed binomial coefficients }are defined by%
\begin{equation}%
%TCIMACRO{\QATOPD{[}{]}{\alpha}{m}}%
%BeginExpansion
\genfrac{[}{]}{0pt}{}{\alpha}{m}%
%EndExpansion
_{q^{a}}\equiv\frac{\left[  \left[  \alpha\right]  \right]  _{q^{a}}\left[
\left[  \alpha-1\right]  \right]  _{q^{a}}\ldots\left[  \left[  \alpha
-m+1\right]  \right]  _{q^{a}}}{\left[  \left[  m\right]  \right]  _{q^{a}}!},
\end{equation}
with $\alpha\in\mathbb{C},$ $m\in\mathbb{N}$. Applying the relations of Eqs.
(\ref{Psi-12dimKoord}) and (\ref{RekBrai}) in succession finally yields%
\begin{align}
&  \Psi_{X,Y}^{-1}\left(  (X^{1})^{n_{1}}(X^{2})^{n_{2}}\otimes(Y^{1})^{m_{1}%
}(Y^{2})^{m_{2}}\right) \label{BraidMon2dim}\\
&  =\sum_{i=0}^{\min(n_{2},m_{1})}(-\lambda)^{i}\,[[i]]_{q^{-2}}!%
%TCIMACRO{\QATOPD{[}{]}{n_{2}}{i}}%
%BeginExpansion
\genfrac{[}{]}{0pt}{}{n_{2}}{i}%
%EndExpansion
_{q^{-2}}%
%TCIMACRO{\QATOPD{[}{]}{m_{1}}{i}}%
%BeginExpansion
\genfrac{[}{]}{0pt}{}{m_{1}}{i}%
%EndExpansion
_{q^{-2}}\,q^{-(n_{2}-\,i)(m_{1}-\,i)}\nonumber\\
&  \times q^{-2(n_{2}-\,i)(m_{2}+i)-2(m_{1}-\,i)(n_{1}+\,i)-2(n_{1}%
+\,i)(m_{2}+i)+i(m_{1}+\,n_{2}-\,i)}\nonumber\\
&  \times(Y^{1})^{m_{1}-\,i}(Y^{2})^{m_{2}+i}\otimes(X^{1})^{n_{1}+i}%
(X^{2})^{n_{2}-\,i}.\nonumber
\end{align}

In order to get a formula for the braiding between arbitrary elements one has
to realize that the monomials $(X^{1})^{n_{1}}(X^{2})^{n_{2}}$ form a basis of
two-dimensional quantum plane. Due to this fact and the linearity of the
mapping $\Psi_{X,Y}^{-1}$, we are nearly done. What remains is to find a
corresponding expression on commutative algebras. As a first step to this aim,
we fix the isomorphism of Eq. (\ref{AlgIso}) by
\begin{equation}
\mathcal{W}\left(  (x^{1})^{n_{1}}(x^{2})^{n_{2}}\right)  \equiv(X^{1}%
)^{n_{1}}(X^{2})^{n_{2}},\quad n_{1},n_{2}\in\mathbb{N}.
\end{equation}
Recalling\ the definition of the braided product, which implies%
\begin{align}
&  (x^{1})^{n_{1}}(x^{2})^{n_{2}}\,\underline{\odot}_{L}\,(y^{1})^{m_{1}%
}(y^{2})^{m_{2}}\\
&  =\left(  \mathcal{W}^{-1}\mathcal{\otimes W}^{-1}\right)  \circ\Psi
_{X,Y}^{-1}\left(  \mathcal{W}\left(  (x^{1})^{n_{1}}(x^{2})^{n_{2}}\right)
\otimes\mathcal{W}\left(  (y^{1})^{m_{1}}(y^{2})^{m_{2}}\right)  \right)
,\nonumber
\end{align}
we can immediately read off from Eq. (\ref{BraidMon2dim}) that%
\begin{align}
&  (x^{1})^{n_{1}}(x^{2})^{n_{2}}\,\underline{\odot}_{L}\,(y^{1})^{m_{1}%
}(y^{2})^{m_{2}}\label{ComBrai2dim}\\
&  =\sum_{i=0}^{\min(n_{2},m_{1})}(-\lambda)^{i}\,[[i]]_{q^{-2}}!%
%TCIMACRO{\QATOPD{[}{]}{n_{2}}{i}}%
%BeginExpansion
\genfrac{[}{]}{0pt}{}{n_{2}}{i}%
%EndExpansion
_{q^{-2}}%
%TCIMACRO{\QATOPD{[}{]}{m_{1}}{i}}%
%BeginExpansion
\genfrac{[}{]}{0pt}{}{m_{1}}{i}%
%EndExpansion
_{q^{-2}}\,q^{\,i^{2}-i(m_{1}+n_{2})-2i(m_{2}+n_{1})}\nonumber\\
&  \times q^{-(n_{2}-\,i)(m_{1}-\,i)-2(n_{2}-\,i)m_{2}-2n_{1}(m_{1}%
-\,i)-2n_{1}m_{2}}\nonumber\\
&  \times(y^{1})^{m_{1}-\,i}(y^{2})^{m_{2}+i}\otimes(x^{1})^{n_{1}+\,i}%
(x^{2})^{n_{2}-\,i}\nonumber\\
&  =\sum_{i=0}^{\min(n_{2},m_{1})}q^{\,i^{2}}(-\lambda)^{i}\,\frac{(y^{2}%
)^{i}\otimes(x^{1})^{i}}{[[i]]_{q^{-2}}!}\,q^{-\hat{n}_{1}\otimes\,\hat{n}%
_{2}-2\hat{n}_{2}\otimes\,\hat{n}_{2}-2\hat{n}_{1}\otimes\,\hat{n}_{1}%
-2\hat{n}_{2}\otimes\,\hat{n}_{1}}\nonumber\\
&  \times(D_{q^{-2}}^{1})^{i}(q^{-i}y^{1})^{m_{1}}\,(q^{-2i}y^{2})^{m_{2}%
}\otimes(q^{-2i}x^{1})^{n_{1}}\,(D_{q^{-2}}^{2})^{i}(q^{-i}x^{2})^{n_{2}%
},\nonumber
\end{align}
where for the second identity we have rewritten the powers of $q$ and the
q-binomial coefficients by making use of%
\begin{equation}
(D_{q^{a}}^{\alpha})^{i}(x^{1})^{n_{1}}(x^{2})^{n_{2}}=[[i]]_{q^{a}}!%
%TCIMACRO{\QATOPD{[}{]}{n_{\alpha}}{i}}%
%BeginExpansion
\genfrac{[}{]}{0pt}{}{n_{\alpha}}{i}%
%EndExpansion
_{q^{a}}(x^{\alpha})^{-i}(x^{1})^{n_{1}}(x^{2})^{n_{2}},\quad\alpha=1,2,
\end{equation}
(to be more precise, we should mention that we have $(D_{q^{a}}^{\alpha}%
)^{i}(x^{1})^{n_{1}}(x^{2})^{n_{2}}=0$ for $i>n_{\alpha}$) and%
\begin{align}
&  q^{a(\hat{n}_{\alpha}\otimes\,\hat{n}_{\beta})}\left(  (y^{1})^{m_{1}%
}(y^{2})^{m_{2}}\otimes(x^{1})^{n_{1}}(x^{2})^{n_{2}}\right) \\
&  =q^{a(m_{\alpha}n_{\beta})}\left(  (y^{1})^{m_{1}}(y^{2})^{m_{2}}%
\otimes(x^{1})^{n_{1}}(x^{2})^{n_{2}}\right)  .\nonumber
\end{align}
Now, it is not very difficult to convince oneself that the last operator
expression in Eq. (\ref{ComBrai2dim}) also holds for arbitrary commutative
functions (at least those which can be expanded in an absolutely convergent
power series). Thus, we end up with the following general expression for the
braided product of two commutative functions:%
\begin{align}
&  f(x^{1},x^{2})\,\underline{\odot}_{L}\,g(y^{1},y^{2})\\
&  =\sum_{i=0}^{\infty}q^{\,i^{2}}(-\lambda)^{i}\,\frac{(y^{2})^{i}%
\otimes(x^{1})^{i}}{[[i]]_{q^{-2}}!}\,q^{-\hat{n}_{1}\otimes\,\hat{n}%
_{2}-2\hat{n}_{2}\otimes\,\hat{n}_{2}-2\hat{n}_{1}\otimes\,\hat{n}_{1}%
-2\hat{n}_{2}\otimes\,\hat{n}_{1}}\nonumber\\
&  \times(D_{q^{-2}}^{1})^{i}g(q^{-i}y^{1},q^{-2i}y^{2})\otimes(D_{q^{-2}}%
^{2})^{i}f(q^{-2i}x^{1},q^{-i}x^{2}).\nonumber
\end{align}

Next, we would like to move on to the corresponding formula concerning the
braiding with respect to $\Psi_{X,Y}$. Starting from the relations in Eq.
(\ref{BraidKoord2dim}) and repeating the same steps as before, one can obtain%
\begin{align}
&  \Psi_{X,Y}\left(  (X^{2})^{n_{2}}(X^{1})^{n_{1}}\otimes(Y^{2})^{m_{2}%
}(Y^{1})^{m_{1}}\right) \\
&  =\sum_{i=0}^{\min(n_{1},m_{2})}\lambda^{i}\,[[i]]_{q^{2}}!%
%TCIMACRO{\QATOPD{[}{]}{n_{1}}{i}}%
%BeginExpansion
\genfrac{[}{]}{0pt}{}{n_{1}}{i}%
%EndExpansion
_{q^{2}}%
%TCIMACRO{\QATOPD{[}{]}{m_{2}}{i}}%
%BeginExpansion
\genfrac{[}{]}{0pt}{}{m_{2}}{i}%
%EndExpansion
_{q^{2}}\,q^{(n_{1}-\,i)(m_{2}-i)}\nonumber\\
&  \times q^{2(n_{1}-\,i)(m_{1}+\,i)+2(m_{2}-\,i)(n_{2}+i)+2(n_{2}%
+i)(m_{1}+\,i)-i(m_{2}+n_{1}-\,i)}\nonumber\\
&  \times(Y^{2})^{m_{2}-\,i}(Y^{1})^{m_{1}+\,i}\otimes(X^{2})^{n_{2}+i}%
(X^{1})^{n_{1}-\,i}.\nonumber
\end{align}
Specifying the algebra isomorphism by
\begin{equation}
\mathcal{W}\left(  (x^{1})^{n_{1}}(x^{2})^{n_{2}}\right)  =(X^{2})^{n_{2}%
}(X^{1})^{n_{1}},\quad n_{1},n_{2}\in\mathbb{N},
\end{equation}
leads us to
\begin{align}
&  f(x^{2},x^{1})\,\underline{\widetilde{\odot}}_{\bar{L}}\,g(y^{2}%
,y^{1})\label{BraidTild}\\
&  =\sum_{i=0}^{\infty}q^{-i^{2}}\lambda^{i}\,\frac{(y^{1})^{i}\otimes
(x^{2})^{i}}{[[i]]_{q^{2}}!}\,q^{\hat{n}_{2}\otimes\hat{n}_{1}+2\hat{n}%
_{1}\otimes\hat{n}_{1}+2\hat{n}_{2}\otimes\hat{n}_{2}+2\hat{n}_{1}\otimes
\hat{n}_{2}}\nonumber\\
&  \times(D_{q^{2}}^{2})^{i}g(q^{i}y^{2},q^{2i}y^{1})\otimes(D_{q^{2}}%
^{1})^{i}f(q^{2i}x^{2},q^{i}x^{1}),\nonumber
\end{align}
showing, in turn, that the following connection between the two braided
products holds:%
\begin{equation}
f(x^{2},x^{1})\,\underline{\widetilde{\odot}}_{\bar{L}}\,g(y^{2}%
,y^{1})\overset{{%
%TCIMACRO{\QATOP{\alpha}{q}}%
%BeginExpansion
\genfrac{}{}{0pt}{}{\alpha}{q}%
%EndExpansion
}{%
%TCIMACRO{\QATOP{\rightarrow}{\rightarrow}}%
%BeginExpansion
\genfrac{}{}{0pt}{}{\rightarrow}{\rightarrow}%
%EndExpansion
}{%
%TCIMACRO{\QATOP{\alpha^{\prime}}{1/q}}%
%BeginExpansion
\genfrac{}{}{0pt}{}{\alpha^{\prime}}{1/q}%
%EndExpansion
}}{\longleftrightarrow}f(x^{1},x^{2})\,\underline{\odot}_{L}\,g(y^{1},y^{2}),
\end{equation}
where the symbol $\overset{{%
%TCIMACRO{\QATOP{\alpha}{q}}%
%BeginExpansion
\genfrac{}{}{0pt}{}{\alpha}{q}%
%EndExpansion
}{%
%TCIMACRO{\QATOP{\rightarrow}{\rightarrow}}%
%BeginExpansion
\genfrac{}{}{0pt}{}{\rightarrow}{\rightarrow}%
%EndExpansion
}{%
%TCIMACRO{\QATOP{\alpha^{\prime}}{1/q}}%
%BeginExpansion
\genfrac{}{}{0pt}{}{\alpha^{\prime}}{1/q}%
%EndExpansion
}}{\longleftrightarrow}$ indicates a transition via the substitutions
\begin{equation}
D_{q^{a}}^{\alpha}\rightarrow D_{q^{-a}}^{\alpha^{\prime}},\quad\hat
{n}^{\alpha}\rightarrow-\hat{n}^{\alpha^{\prime}}\quad x^{\alpha}\rightarrow
x^{\alpha^{\prime}},\quad q\rightarrow q^{-1}.
\end{equation}
It should be obvious that this correspondence is a direct consequence of the
crossing symmetry in Eq. (\ref{TranRul2dim}). One should also notice that in
Eq. (\ref{BraidTild}) the tilde on the multiplication symbol shall remind us
of the fact that the explicit form of this braided product refers to a
reversed normal ordering.

\section{q-Deformed Euclidean space in three dimensions\label{Kap2}}

In this section we would like to deal with three-dimensional q-deformed
Euclidean space. Its algebra is spanned by three coordinates $X^{A},$
$A\in\{+,3,-\},$ which fulfill as commutation relations \cite{LWW97}
\begin{align}
X^{3}X^{\pm}  &  =q^{\pm2}X^{+}X^{3},\\
X^{-}X^{+}  &  =X^{+}X^{-}+\lambda X^{3}X^{3}.\nonumber
\end{align}
Let us recall that the entries $\mathcal{L}_{B}^{A}$ of the so-called L-matrix
determine the braiding of the quantum space coordinates $X^{A},$
$A\in\{+,3,-\}$, since we have (if not stated otherwise summation over
repeated indices is to be understood)%
\begin{equation}
\Psi_{X,Y}(X^{A}\otimes w)=(\mathcal{L}_{B}^{A}\triangleright w)\otimes X^{B}.
\end{equation}
From the results in Ref. \cite{BW01} we can read off their explicit form in
terms of $U_{q}(su_{2})$-generators, leaving us with%
\begin{align}
\Psi_{X,Y}(X^{+}\otimes w)  &  =(\Lambda^{1/2}(\tau^{3})^{-1/2}\triangleright
w)\otimes X^{+},\label{BraiKoord3dim}\\
\Psi_{X,Y}(X^{3}\otimes w)  &  =(\Lambda^{1/2}\triangleright w)\otimes
X^{3}+\lambda\lambda_{+}(\Lambda^{1/2}L^{-}\triangleright w)\otimes
X^{+},\nonumber\\
\Psi_{X,Y}(X^{-}\otimes w)  &  =(\Lambda^{1/2}(\tau^{3})^{1/2}\triangleright
w)\otimes X^{-}\nonumber\\
&  +\,q^{-1}\lambda\lambda_{+}(\Lambda^{1/2}(\tau^{3})^{1/2}L^{-}%
\triangleright w)\otimes X^{3}\\
&  +\,q^{-2}\lambda^{2}\lambda_{+}(\Lambda^{1/2}(\tau^{3})^{1/2}(L^{-}%
)^{2}\triangleright w)\otimes X^{+},\nonumber
\end{align}
where $w$ denotes an arbitrary element of a quantum space the algebra
$U_{q}(su_{2})$ acts upon. One should notice that $\Lambda$ stands for a
scaling operator satisfying the relations%
\begin{equation}
\Lambda X^{A}=q^{4}X^{A}\Lambda,\qquad\Lambda\triangleright1=1.
\end{equation}

As a next step let us try to derive formulae for the expressions
\begin{align}
&  \Psi_{X,Y}((X^{A})^{n}\otimes w)\\
&  =(\mathcal{L}_{B_{1}}^{A}\mathcal{L}_{B_{2}}^{A}\ldots\mathcal{L}_{B_{n}%
}^{A}\triangleright w)\otimes X^{B_{1}}X^{B_{2}}\ldots X^{B_{n}},\nonumber
\end{align}
where $A\in\{+,3,-\}$ and $n\in\mathbb{N}$. In order to do so, we perform the
following calculation:%
\begin{align}
&  \mathcal{L}_{B_{1}}^{+}\mathcal{L}_{B_{2}}^{+}\ldots\mathcal{L}_{B_{n}}%
^{+}\otimes X^{B_{1}}X^{B_{2}}\ldots X^{B_{n}}=(\mathcal{L}_{B}^{+}\otimes
X^{B})^{n}\label{BraidXplus}\\
&  =(\Lambda^{1/2}(\tau^{3})^{-1/2}\otimes X^{+})^{n}=(\Lambda^{1/2}(\tau
^{3})^{-1/2})^{n}\otimes(X^{+})^{n}.\nonumber
\end{align}
Proceeding in the same way for $X^{3}$ and recalling the q-binomial theorem
\cite{Koo96, KS97} for q-commuting variables, i.e.%
\begin{equation}
(A+B)^{n}=\sum_{k=0}^{n}%
%TCIMACRO{\QATOPD{[}{]}{n}{k}}%
%BeginExpansion
\genfrac{[}{]}{0pt}{}{n}{k}%
%EndExpansion
_{q^{a}}A^{k}B^{n-k},\quad\text{if~}BA=q^{a}AB,
\end{equation}
we obtain
\begin{align}
&  \mathcal{L}_{B_{1}}^{3}\mathcal{L}_{B_{2}}^{3}\ldots\mathcal{L}_{B_{n}}%
^{3}\otimes X^{B_{1}}X^{B_{2}}\ldots X^{B_{n}}=(\mathcal{L}_{B}^{3}\otimes
X^{B})^{n}\label{BraidX3}\\
&  =(\Lambda^{1/2}\otimes X^{3}+\lambda\lambda_{+}\Lambda^{1/2}L^{-}\otimes
X^{+})^{n}\nonumber\\
&  =\sum_{k=0}^{n}%
%TCIMACRO{\QATOPD{[}{]}{n}{k}}%
%BeginExpansion
\genfrac{[}{]}{0pt}{}{n}{k}%
%EndExpansion
_{q^{2}}(\lambda\lambda_{+})^{k}(\Lambda^{1/2})^{n}(L^{-})^{k}\otimes
(X^{+})^{k}(X^{3})^{n-k}.
\end{align}
Notice that application of the q-binomial theorem in the third identity
requires to hold
\begin{equation}
(\Lambda^{1/2}\otimes X^{3})(\Lambda^{1/2}L^{-}\otimes X^{+})=q^{2}\Lambda
L^{-}\otimes X^{+}X^{3},
\end{equation}
which follows directly from the commutation relations for coordinates and
symmetry generators (see for example Ref. \cite{BW01}).

To derive a relation describing the braiding of powers of $X^{-}$ is a little
bit more involved. It turns out to be convenient to formulate the braiding in
terms of the coproduct. On quantum space coordinates the coproduct takes on
the form \cite{Wac04}, \cite{Mik04}%
\begin{align}
\Delta_{\bar{L}}(X^{A})  &  =(X^{A})_{(1)}\otimes(X^{A})_{(2)}\\
&  =X^{A}\otimes1+\mathcal{L}_{B}^{A}\otimes X^{B}.\nonumber
\end{align}
Now, we are in a position to introduce the identity%
\begin{equation}
\Psi_{X,Y}\left(  (X^{A})^{n}\otimes w\right)  =\left(  \left[  (X^{A}%
)^{n}\right]  _{(1)}\triangleright w\right)  \otimes\left[  (X^{A}%
)^{n}\right]  _{(2)}, \label{BraCo3dim}%
\end{equation}
which is valid, if we demand for quantum space coordinates to act on elements
of another quantum space by%
\begin{equation}
X^{A}\triangleright w=0. \label{ActXX}%
\end{equation}

Next, we would like to derive an explicit formula for the coproduct of powers
of $X^{-}$. To achieve this, we take into account the module coalgebra
property of our quantum spaces \cite{Maj95}, which implies
\begin{align}
&  \Delta_{\bar{L}}\left(  (L^{-})^{n}\triangleright(X^{3})^{n}\right)
=\Delta_{\bar{L}}\left(  \left[  (L^{-})^{n}\right]  _{(1)}(X^{3}%
)^{n}\,S\big(\left[  (L^{-})^{n}\right]  _{(2)}\big)\right)  \label{ComAlgPro}%
\\
&  =\sum_{k=0}^{n}(-1)^{k}q^{k(k-1)}%
%TCIMACRO{\QATOPD{[}{]}{n}{k}}%
%BeginExpansion
\genfrac{[}{]}{0pt}{}{n}{k}%
%EndExpansion
_{q^{2}}\Delta((L^{-})^{n-\,k})\,\Delta_{\bar{L}}((X^{3})^{n})\,\Delta
((L^{-})^{k})\,\Delta((\tau^{3})^{(n-\,k)/2}).\nonumber
\end{align}
In this derivation, the first step uses the explicit form for the adjoint
action and the second equality makes use of
\begin{align}
&  \Delta(L^{-})^{n}=(\Delta(L^{-}))^{n}\label{CoProLmin}\\
&  =(L^{-}\otimes(\tau^{3})^{-1/2}+1\otimes L^{-})^{n}\nonumber\\
&  =\sum_{k=0}^{n}q^{-2k(n-\,k)}%
%TCIMACRO{\QATOPD{[}{]}{n}{k}}%
%BeginExpansion
\genfrac{[}{]}{0pt}{}{n}{k}%
%EndExpansion
_{q^{2}}(L^{-})^{n-\,k}\otimes(L^{-})^{k}(\tau^{3})^{-(n-\,k)/2},\nonumber
\end{align}
together with%
\begin{align}
&  S((L^{-})^{n})=(S(L^{-}))^{n}\\
&  =(-L^{-}(\tau^{3})^{-1/2})^{n}=(-1)^{n}q^{n(n-1)}(L^{-})^{n}((\tau
^{3})^{-1/2})^{n}.\nonumber
\end{align}
On the other hand, using the operator representation of $L^{-}$, as it was
presented in Ref. \cite{BW01}, we get
\begin{equation}
\Delta_{\bar{L}}\left(  (L^{-})^{n}\triangleright(X^{3})^{n}\right)
=q^{-n^{2}}[[n]]_{q^{2}}!\,\Delta_{\bar{L}}\left(  (X^{-})^{n}\right)  .
\label{DarXmin}%
\end{equation}
A short glance at Eqs. (\ref{ComAlgPro}) and (\ref{DarXmin}) shows us how to
reduce coproducts for powers of $X^{-}$ to those for powers of $L^{-},$
$\tau^{3},$ and $X^{3}$. Thus, it remains to find an explicit formula for
coproducts of powers of $X^{3}$ ($\tau^{3}$ is a grouplike element). In Ref.
\cite{Wac04} this missing link was found to be%
\begin{equation}
\Delta_{\bar{L}}(X^{3})^{n}=\sum_{k=0}^{n}%
%TCIMACRO{\QATOPD{[}{]}{n}{k}}%
%BeginExpansion
\genfrac{[}{]}{0pt}{}{n}{k}%
%EndExpansion
_{q^{2}}(\lambda\lambda_{+})^{k}(\Lambda^{1/2})^{n}(L^{-})^{k}\otimes
(X^{+})^{k}(X^{3})^{n-k}+\ldots,
\end{equation}
where the insertion points indicate terms that in the end will not contribute
in Eq. (\ref{BraCo3dim}) due to Eq. (\ref{ActXX}).

From what we have done so far, it is now rather straightforward but laborious
to obtain%
\begin{align}
&  \Psi_{X,Y}((X^{-})^{n}\otimes w)=\left(  \mathcal{L}_{B_{1}}^{-}%
\mathcal{L}_{B_{2}}^{-}\ldots\mathcal{L}_{B_{n}}^{-}\triangleright w\right)
\otimes X^{B_{1}}X^{B_{2}}\ldots X^{B_{n}}\label{BraidXmin}\\
&  =\sum_{i=0}^{n}\sum_{k=0}^{n}\sum_{l=0}^{n-k}(-1)^{k}\,\frac{(\lambda
\lambda_{+})^{i}}{[[n]]_{q^{2}}!}\,q^{n^{2}+\,k(k-1)-2l(n-\,i-k-l)}\nonumber\\
&  \times%
%TCIMACRO{\QATOPD{[}{]}{n}{k}}%
%BeginExpansion
\genfrac{[}{]}{0pt}{}{n}{k}%
%EndExpansion
_{q^{2}}%
%TCIMACRO{\QATOPD{[}{]}{n-k}{l}}%
%BeginExpansion
\genfrac{[}{]}{0pt}{}{n-k}{l}%
%EndExpansion
_{q^{2}}%
%TCIMACRO{\QATOPD{[}{]}{n}{i}}%
%BeginExpansion
\genfrac{[}{]}{0pt}{}{n}{i}%
%EndExpansion
_{q^{2}}\,\Big [(\Lambda^{1/2})^{n}(L^{-})^{i+k+l}(\tau^{3})^{n/2}%
\triangleright w\Big ]\nonumber\\
&  \otimes\,(L^{-})^{n-\,k-l}\triangleright(X^{+})^{i}(X^{3})^{n-\,i}%
.\nonumber
\end{align}
Applying Eqs. (\ref{BraidXplus}), (\ref{BraidX3}), and (\ref{BraidXmin}) in
succession finally yields
\begin{align}
&  \Psi_{X,Y}((X^{+})^{n_{+}}(X^{3})^{n_{3}}(X^{-})^{n_{-}}\otimes
w)\label{BraidKordAlg3dim}\\
&  =(\mathcal{L}_{A_{1}}^{+}\ldots\mathcal{L}_{A_{n_{+}}}^{+}\,\mathcal{L}%
_{B_{1}}^{3}\ldots\mathcal{L}_{B_{n_{3}}}^{3}\,\mathcal{L}_{C_{1}}^{-}%
\ldots\mathcal{L}_{C_{n_{-}}}^{-}\triangleright w)\nonumber\\
&  \otimes X^{A_{1}}\ldots X^{A_{n_{+}}}\,X^{B_{1}}\ldots X^{B_{n_{3}}%
}\,X^{C_{1}}\ldots X^{C_{n_{-}}}\nonumber\\
&  =\sum_{i=0}^{n_{3}}\sum_{j=0}^{n_{-}}\sum_{k=0}^{n_{-}}\sum_{l=0}%
^{n_{-}-\,k}(-1)^{k}\frac{(\lambda\lambda_{+})^{i+j}}{[[n_{-}]]_{q^{2}}%
!}\nonumber\\
&  \times\,q^{k(k-1)+2l(k+l+j)-2n_{+}(i+j+k+l)+n_{-}(n_{-}-2l)}\nonumber\\
&  \times\,%
%TCIMACRO{\QATOPD{[}{]}{k+l}{k}}%
%BeginExpansion
\genfrac{[}{]}{0pt}{}{k+l}{k}%
%EndExpansion
_{q^{2}}%
%TCIMACRO{\QATOPD{[}{]}{n_{3}}{i}}%
%BeginExpansion
\genfrac{[}{]}{0pt}{}{n_{3}}{i}%
%EndExpansion
_{q^{2}}%
%TCIMACRO{\QATOPD{[}{]}{n_{-}}{j}}%
%BeginExpansion
\genfrac{[}{]}{0pt}{}{n_{-}}{j}%
%EndExpansion
_{q^{2}}%
%TCIMACRO{\QATOPD{[}{]}{n_{-}}{k+l}}%
%BeginExpansion
\genfrac{[}{]}{0pt}{}{n_{-}}{k+l}%
%EndExpansion
_{q^{2}}\nonumber\\
&  \times\,\Big [(\Lambda^{1/2})^{n_{+}+\,n_{3}+n_{-}}(L^{-})^{i+j+k+l}%
(\tau^{3})^{(n_{-}-\,n_{+})/2}\triangleright w\Big ]\nonumber\\
&  \otimes\,(X^{+})^{n_{+}+\,i}(X^{3})^{n_{3}-i}\Big [(L^{-})^{n_{-}%
-\,k-l}\triangleright(X^{+})^{j}(X^{3})^{n_{-}-\,j}\Big ].\nonumber
\end{align}

There remains to evaluate the action in the last expression of Eq.
(\ref{BraidKordAlg3dim}). This can be achieved by formula (\ref{WirkLMon3dim})
in Appendix \ref{ActSymAlg}. Now, we have everything together to write down an
expression for the braiding of two normal ordered monomials. Having finished
that task, we can follow the same line of arguments as in the previous
section. More concretely, we determine the algebra isomorphism in Eq.
(\ref{AlgIso}) by
\begin{equation}
\mathcal{W}\left(  (x^{+})^{n_{+}}(x^{3})^{n_{3}}(x^{-})^{n_{-}}\right)
\equiv(X^{+})^{n_{+}}(X^{3})^{n_{3}}(X^{-})^{n_{-}},\quad n_{+},n_{3},n_{-}%
\in\mathbb{N},
\end{equation}
and read off from (\ref{BraidKordAlg3dim}) an expression for the braided
product in very much the same way as was done for the two-dimensional case. In
this manner, we finally arrive at (if not stated otherwise, all summation
variables in this article have to be integer and non-negative)
\begin{align}
&  f(x^{+},x^{3},x^{-})\,\underline{\odot}_{\bar{L}}\,g(y^{+},y^{3},y^{-})\\
&  =\sum_{i,s=0}^{\infty}\sum_{j=0}^{s}\sum_{k+l+t=s}\sum_{{%
%TCIMACRO{\QATOP{u+v=t}{0\leq v\leq u\leq j}}%
%BeginExpansion
\genfrac{}{}{0pt}{}{u+v=t}{0\leq v\leq u\leq j}%
%EndExpansion
}}(-1)^{k}\,(\lambda\lambda_{+})^{i+j}\,t_{u,v}\nonumber\\
&  \times\,q^{-k-v+2(k+l+i+j)^{2}+2v(s-v)+2u(i-v)-2j(i-l)+\,l^{2}+lt+t^{2}%
}\nonumber\\
&  \times\frac{\lbrack\lbrack u]]_{q^{4}}!}{[[i]]_{q^{4}}![[k]]_{q^{4}%
}![[l]]_{q^{4}}![[s]]_{q^{4}}![[t]]_{q^{4}}!}%
%TCIMACRO{\QATOPD{[}{]}{j}{u}}%
%BeginExpansion
\genfrac{[}{]}{0pt}{}{j}{u}%
%EndExpansion
_{q^{4}}%
%TCIMACRO{\QATOPD{[}{]}{s}{j}}%
%BeginExpansion
\genfrac{[}{]}{0pt}{}{s}{j}%
%EndExpansion
_{q^{2}}\nonumber\\
&  \times\,q^{2(\hat{n}_{+}+\,\hat{n}_{3}+\,\hat{n}_{-})\otimes(\hat{n}%
_{+}+\,\hat{n}_{3}+\,\hat{n}_{-})+2(\hat{n}_{+}-\,\hat{n}_{-})\otimes(\hat
{n}_{+}-\,\hat{n}_{-})}\nonumber\\
&  \times\,\Big [(L^{-})^{i+j+k+l}\triangleright g(q^{-2(k+l+i+j)}%
y^{+},q^{2(k+l+i+j)}y^{-})\nonumber\\
&  \,\quad\otimes(x^{+})^{i+j-u}(x^{3})^{s-j+u-v}(x^{-})^{v}\nonumber\\
&  \,\quad\times(D_{q^{2}}^{3})^{i}(D_{q^{2}}^{-})^{s}f(q^{2(j-u)}%
x^{3},q^{-2(i+j+l)}x^{-})\Big ],\nonumber
\end{align}
where the explicit form of $t_{u,v}$ and that for the action of powers of
$L^{-}$ can again be looked up in Appendix \ref{ActSymAlg} [see Eqs.
(\ref{tuv}) and (\ref{WirL-Func})].

Let us end with the comment that we could also have started our considerations
from the other braiding determined by%
\begin{align}
\Psi_{X,Y}^{-1}(X^{-}\otimes w)  &  =(\Lambda^{-1/2}(\tau^{3})^{-1/2}%
\triangleright w)\otimes X^{-},\\
\Psi_{X,Y}^{-1}(X^{3}\otimes w)  &  =(\Lambda^{-1/2}\triangleright w)\otimes
X^{3}+\lambda\lambda_{+}(\Lambda^{-1/2}L^{+}\triangleright w)\otimes
X^{-},\nonumber\\
\Psi_{X,Y}^{-1}(X^{+}\otimes w)  &  =(\Lambda^{-1/2}(\tau^{3})^{1/2}%
\triangleright w)\otimes X\nonumber\\
&  +q\lambda\lambda_{+}(\Lambda^{-1/2}(\tau^{3})^{1/2}L^{+}\triangleright
w)\otimes X^{3}\\
&  +\,q^{2}\lambda^{2}\lambda_{+}(\Lambda^{-1/2}(\tau^{3})^{1/2}(L^{+}%
)^{2}\triangleright w)\otimes X^{-}.\nonumber
\end{align}
This way we would get a second braided product denoted by $\underline
{\widetilde{\odot}}_{L},$ which is linked to the first one via the
transformation rule%
\begin{equation}
f(x^{+},x^{3},x^{-})\,\underline{\odot}_{\bar{L}}\,g(y^{+},y^{3}%
,y^{-})\overset{{%
%TCIMACRO{\QATOP{\pm}{q}}%
%BeginExpansion
\genfrac{}{}{0pt}{}{\pm}{q}%
%EndExpansion
}{%
%TCIMACRO{\QATOP{\rightarrow}{\rightarrow}}%
%BeginExpansion
\genfrac{}{}{0pt}{}{\rightarrow}{\rightarrow}%
%EndExpansion
}{%
%TCIMACRO{\QATOP{\mp}{1/q}}%
%BeginExpansion
\genfrac{}{}{0pt}{}{\mp}{1/q}%
%EndExpansion
}}{\longleftrightarrow}f(x^{-},x^{3},x^{+})\,\underline{\widetilde{\odot}}%
_{L}\,g(y^{-},y^{3},y^{+}),
\end{equation}
where $\overset{{%
%TCIMACRO{\QATOP{\pm}{q}}%
%BeginExpansion
\genfrac{}{}{0pt}{}{\pm}{q}%
%EndExpansion
}{%
%TCIMACRO{\QATOP{\rightarrow}{\rightarrow}}%
%BeginExpansion
\genfrac{}{}{0pt}{}{\rightarrow}{\rightarrow}%
%EndExpansion
}{%
%TCIMACRO{\QATOP{\mp}{1/q}}%
%BeginExpansion
\genfrac{}{}{0pt}{}{\mp}{1/q}%
%EndExpansion
}}{\longleftrightarrow}$ indicates that we can make a transition between the
two expressions by applying the substitutions%
\begin{equation}
D_{q^{a}}^{\pm}\rightarrow D_{q^{-a}}^{\mp},\quad\hat{n}_{\pm}\rightarrow
-\hat{n}_{\mp},\quad x^{\pm}\rightarrow x^{\mp},\quad q^{\pm1}\rightarrow
q^{\mp1}.
\end{equation}

\section{\label{Kap3}q-Deformed Euclidean space in four dimensions}

The four-dimensional Euclidean space \cite{Oca96} can be treated in very much
the same way as the three-dimensional one. Therefore we will restrict
ourselves to stating the results, only. Again, we begin by considering its
defining relation, which in terms of the coordinates $X^{i},$ $i\in
\{1,\ldots,4\}$, explicitly read
\begin{align}
X^{1}X^{j}  &  =qX^{j}X^{1},\\
X^{j}X^{4}  &  =qX^{4}X^{j},\quad j=2,3,\nonumber\\
X^{2}X^{3}  &  =X^{3}X^{2},\nonumber\\
X^{4}X^{1}  &  =X^{1}X^{4}+\lambda X^{2}X^{3}.\nonumber
\end{align}

The results in Ref. \cite{BW01} tell us that the inverse braiding of a single
quantum space coordinate with an element $w$ of another quantum space is now
determined by%
\begin{align}
\Psi_{X,Y}^{-1}(X^{1}\otimes w)  &  =(\Lambda^{-1/2}K_{1}^{1/2}K_{2}%
^{1/2}\triangleright w)\otimes X^{1},\label{BraidKoord4dim}\\
\Psi_{X,Y}^{-1}(X^{2}\otimes w)  &  =(\Lambda^{-1/2}K_{1}^{-1/2}K_{2}%
^{1/2}\triangleright w)\otimes X^{2}\nonumber\\
&  +q\lambda(\Lambda^{-1/2}K_{1}^{1/2}K_{2}^{1/2}L_{1}^{+}\triangleright
w)\otimes X^{1},\nonumber\\
\Psi_{X,Y}^{-1}(X^{3}\otimes w)  &  =(\Lambda^{-1/2}K_{1}^{1/2}K_{2}%
^{-1/2}\triangleright w)\otimes X^{3}\nonumber\\
&  +q\lambda(\Lambda^{-1/2}K_{1}^{1/2}K_{2}^{1/2}L_{2}^{+}\triangleright
w)\otimes X^{1},\nonumber\\
\Psi_{X,Y}^{-1}(X^{4}\otimes w)  &  =(\Lambda^{-1/2}K_{1}^{-1/2}K_{2}%
^{-1/2}\triangleright w)\otimes X^{4}\nonumber\\
&  -\,q\lambda(\Lambda^{-1/2}K_{1}^{-1/2}K_{2}^{1/2}L_{2}^{+}\triangleright
w)\otimes X^{2},\nonumber\\
&  -\,q\lambda(\Lambda^{-1/2}K_{1}^{1/2}K_{2}^{-1/2}L_{1}^{+}\triangleright
w)\otimes X^{3}\nonumber\\
&  -\,q^{2}\lambda^{2}(\Lambda^{-1/2}K_{1}^{1/2}K_{2}^{1/2}L_{1}^{+}L_{2}%
^{+}\triangleright w)\otimes X^{1},\nonumber
\end{align}
where the symmetry generators are part of $U_{q}(so_{4})$ (for its defining
relations see again Appendix \ref{QuanAlg}). With the same reasonings as in
the previous section the identities in Eq. (\ref{BraidKoord4dim}) now imply%
\begin{align}
&  \Psi_{X,Y}^{-1}((X^{1})^{n_{1}}\otimes w)\\
&  =\left(  (\Lambda^{-1/2}K_{1}^{1/2}K_{2}^{1/2})^{n_{1}}\triangleright
w\right)  \otimes(X^{1})^{n_{1}},\nonumber\\[0.16in]
&  \Psi_{X,Y}^{-1}((X^{2})^{n_{2}}\otimes w)\\
&  =\sum_{k=0}^{n_{2}}%
%TCIMACRO{\QATOPD{[}{]}{n_{2}}{k}}%
%BeginExpansion
\genfrac{[}{]}{0pt}{}{n_{2}}{k}%
%EndExpansion
_{q^{-2}}(q\lambda)^{i}\left(  (\Lambda^{-1/2})^{n_{2}}(K_{1}^{1/2}K_{2}%
^{1/2}L_{1}^{+})^{k}(K_{1}^{-1/2}K_{2}^{1/2})^{n_{2}-\,i}\triangleright
w\right) \nonumber\\
&  \,\quad\otimes(X^{1})^{k}(X^{2})^{n_{2}-\,k},\nonumber\\[0.16in]
&  \Psi_{X,Y}^{-1}((X^{3})^{n_{3}}\otimes w)\\
&  =\sum_{k=0}^{n_{3}}%
%TCIMACRO{\QATOPD{[}{]}{n_{3}}{k}}%
%BeginExpansion
\genfrac{[}{]}{0pt}{}{n_{3}}{k}%
%EndExpansion
_{q^{-2}}(q\lambda)^{i}\left(  (\Lambda^{-1/2})^{n_{3}}(K_{1}^{1/2}K_{2}%
^{1/2}L_{2}^{+})^{k}(K_{1}^{1/2}K_{2}^{-1/2})^{n_{3}-\,i}\triangleright
w\right) \nonumber\\
&  \,\quad\otimes(X^{1})^{k}(X^{3})^{n_{3}-\,k}.\nonumber
\end{align}
The corresponding formula for the quantum space coordinate $X^{4}$ takes on a
much more complicated form, i.e.
\begin{align}
&  \Psi_{X,Y}^{-1}((X^{4})^{n_{1}}\otimes w)\label{BraiX44dim}\\
&  =\sum_{k=0}^{n_{4}}\sum_{u=0}^{n_{4}}\sum_{v=0}^{n_{4}-k}(-1)^{k}%
\frac{(q\lambda)^{u}}{[[n_{4}]]_{q^{2}}!}q^{n_{4}(n_{4}%
+1)/2+k(k+1)-u(u-1)/2-n_{4}(k+v)}\nonumber\\
&  \times\,q^{2v(n_{4}-k-v)+u(n_{4}-u)}%
%TCIMACRO{\QATOPD{[}{]}{n_{4}}{k}}%
%BeginExpansion
\genfrac{[}{]}{0pt}{}{n_{4}}{k}%
%EndExpansion
_{q^{2}}%
%TCIMACRO{\QATOPD{[}{]}{n_{4}-k}{v}}%
%BeginExpansion
\genfrac{[}{]}{0pt}{}{n_{4}-k}{v}%
%EndExpansion
_{q^{-2}}%
%TCIMACRO{\QATOPD{[}{]}{n_{4}}{u}}%
%BeginExpansion
\genfrac{[}{]}{0pt}{}{n_{4}}{u}%
%EndExpansion
_{q^{-2}}\nonumber\\
&  \times\,\left(  (\Lambda^{-1/2})^{n_{4}}(K_{1}^{1/2})^{-n_{4}+2(k+v)}%
(K_{2}^{1/2})^{-n_{4}+2u}(L_{1}^{+})^{k+v}(L_{2}^{+})^{u}\right)
\triangleright w\nonumber\\
&  \otimes\,(L_{1}^{+})^{n_{4}-k-v}\triangleright(X^{1})^{u}(X^{3})^{n_{4}%
-u}.\nonumber
\end{align}
The derivation of this relation is mainly based on
\begin{align}
&  \Delta_{L}\left(  (L_{1}^{+})^{n}\triangleright(X^{3})^{n}\right)
=\Delta_{L}\left(  \left[  (L_{1}^{+})^{n}\right]  _{(1)}(X^{3})^{n}%
\,S\big(\left[  (L_{1}^{+})^{n}\right]  _{(2)}\big)\right) \\
&  =\sum_{k=0}^{n}(-1)^{n-k}\,q^{(k+1)(n-k)}%
%TCIMACRO{\QATOPD{[}{]}{n}{k}}%
%BeginExpansion
\genfrac{[}{]}{0pt}{}{n}{k}%
%EndExpansion
_{q^{-2}}\Delta((L_{1}^{+})^{\,k})\,\Delta_{L}((X^{3})^{n})\,\Delta((L_{1}%
^{+})^{n-k})\nonumber
\end{align}
and
\begin{equation}
\Delta_{L}\left(  (L_{1}^{+})^{n}\triangleright(X^{3})^{n}\right)
=q^{-n(n+1)/2}[[n]]_{q^{2}}!\,\Delta_{L}(X^{4})^{n},
\end{equation}
since both identities allow one to proceed in a similar fashion as for the
coordinate $X^{-}$ of three-dimensional q-deformed Euclidean space. A short
glance at Eq. (\ref{BraiX44dim}) shows us that we also need to know the action
of powers of $L_{i}^{+},$ $i=1,2,$ on normal ordered monomials (the action of
the $K_{i},$ $i=1,2$, is very simple). Explicit formulae for this task have
been derived in Appendix \ref{ActSymAlg} [see Eqs. (\ref{L1+Wirk}) and
(\ref{L2+Wirk})].

Now, we have everything together for calculating the braiding between a normal
ordered monomial and an arbitrary element of another quantum space. Before
doing so, we fix the algebra isomorphism of Eq. (\ref{AlgIso}) by
\begin{equation}
\mathcal{W}\left(  (x^{1})^{n_{1}}(x^{2})^{n_{2}}(x^{3})^{n_{3}}(x^{4}%
)^{n_{4}}\right)  \equiv(X^{1})^{n_{1}}(X^{2})^{n_{2}}(X^{3})^{n_{3}}%
(X^{4})^{n_{4}},\quad n_{i}\in\mathbb{N}.
\end{equation}
With this convention we finally arrive at%
\begin{align}
&  f(x^{1},x^{2},x^{3},x^{4})\,\underline{\odot}_{L}\,g(y^{1},y^{2}%
,y^{3},y^{4})\\
&  =\sum_{i,j=0}^{\infty}\sum_{s=0}^{\infty}\sum_{%
%TCIMACRO{\QATOP{k+l+v=s}{0\leq v\leq u\leq s}}%
%BeginExpansion
\genfrac{}{}{0pt}{}{k+l+v=s}{0\leq v\leq u\leq s}%
%EndExpansion
}(-1)^{k+v}(q\lambda)^{i+j+u}q^{i(i+1)/2+j(j+1)/2+k(k+1)}\nonumber\\
&  \times\,q^{u(u+1)/2+v(v-1)/2+s(s+1)/2+(u-v)(i+j-2s)-u(v+2j)+i(j-2k-2l)}%
\nonumber\\
&  \times\,\frac{1}{[[i]]_{q^{2}}![[j]]_{q^{2}}![[k]]_{q^{2}}![[l]]_{q^{2}}!}%
%TCIMACRO{\QATOPD{[}{]}{u}{v}}%
%BeginExpansion
\genfrac{[}{]}{0pt}{}{u}{v}%
%EndExpansion
_{q^{2}}%
%TCIMACRO{\QATOPD{[}{]}{s}{u}}%
%BeginExpansion
\genfrac{[}{]}{0pt}{}{s}{u}%
%EndExpansion
_{q^{2}}\nonumber\\
&  \times\,q^{(\hat{n}_{1}+\hat{n}_{2}+\hat{n}_{3}+\hat{n}_{4})\otimes(\hat
{n}_{1}+\hat{n}_{2}+\hat{n}_{3}+\hat{n}_{4})-(\hat{n}_{1}-\,\hat{n}%
_{4})\otimes(\hat{n}_{1}-\,\hat{n}_{4})-(\hat{n}_{2}-\,\hat{n}_{3}%
)\otimes(\hat{n}_{2}-\,\hat{n}_{3})}\nonumber\\
&  \times\,\Big [(L_{1}^{+})^{i+l+k}(L_{2}^{+})^{j+u}\triangleright
g(y^{1},y^{2},y^{3},y^{4})\nonumber\\
&  \,\quad\otimes(x^{1})^{i+j+u-v}(x^{2})^{v}(x^{3})^{s-u}\nonumber\\
&  \,\quad\times(D_{q^{2}}^{2})^{i}(D_{q^{2}}^{3})^{j}(D_{q^{2}}^{4}%
)^{s}f(q^{-(i+j+u-v)}x^{2},q^{-(i+j+u-v)}x^{2},q^{i+j+u-v-s}x^{4}%
)\Big ],\nonumber
\end{align}
where formulae for the action of powers of $L_{i}^{+},$ $i=1,2,$ on
commutative functions are again given in Appendix \ref{ActSymAlg} [see Eqs.
(\ref{ActfL1+}) and (\ref{ActfL2+})].

In complete analogy to the three-dimensional case we can also assign a braided
product to the braiding determined by%
\begin{align}
\Psi_{X,Y}(X^{4}\otimes w)  &  =(\Lambda^{1/2}K_{1}^{1/2}K_{2}^{1/2}%
\triangleright w)\otimes X^{4},\\
\Psi_{X,Y}(X^{3}\otimes w)  &  =(\Lambda^{1/2}K_{1}^{-1/2}K_{2}^{1/2}%
\triangleright w)\otimes X^{3}\nonumber\\
&  +q^{-1}\lambda(\Lambda^{1/2}K_{1}^{1/2}K_{2}^{1/2}L_{1}^{-}\triangleright
w)\otimes X^{4},\nonumber\\
\Psi_{X,Y}(X^{2}\otimes w)  &  =(\Lambda^{1/2}K_{1}^{1/2}K_{2}^{-1/2}%
\triangleright w)\otimes X^{2}\nonumber\\
&  +q^{-1}\lambda(\Lambda^{1/2}K_{1}^{1/2}K_{2}^{1/2}L_{2}^{-}\triangleright
w)\otimes X^{4},\nonumber\\
\Psi_{X,Y}(X^{1}\otimes w)  &  =(\Lambda^{1/2}K_{1}^{-1/2}K_{2}^{-1/2}%
\triangleright w)\otimes X^{1}\nonumber\\
&  -\,q^{-1}\lambda(\Lambda^{1/2}K_{1}^{1/2}K_{2}^{-1/2}L_{1}^{-}%
\triangleright w)\otimes X^{2}\nonumber\\
&  -\,q^{-1}\lambda(\Lambda^{1/2}K_{1}^{-1/2}K_{2}^{1/2}L_{2}^{-}%
\triangleright w)\otimes X^{3}\nonumber\\
&  -\,q^{-2}\lambda^{2}(\Lambda^{1/2}K_{1}^{1/2}K_{2}^{1/2}L_{1}^{-}L_{2}%
^{-}\triangleright w)\otimes X^{4}.\nonumber
\end{align}
The explicit form of that second braided product is related to the first one
by the correspondence
\begin{align}
&  f(x^{1},x^{2},x^{3},x^{4})\,\underline{\odot}_{L}\,g(y^{1},y^{2}%
,y^{3},y^{4})\\
\overset{{{%
%TCIMACRO{\QATOP{i}{q}}%
%BeginExpansion
\genfrac{}{}{0pt}{}{i}{q}%
%EndExpansion
}{%
%TCIMACRO{\QATOP{\rightarrow}{\rightarrow}}%
%BeginExpansion
\genfrac{}{}{0pt}{}{\rightarrow}{\rightarrow}%
%EndExpansion
}{%
%TCIMACRO{\QATOP{i^{\prime}}{1/q}}%
%BeginExpansion
\genfrac{}{}{0pt}{}{i^{\prime}}{1/q}%
%EndExpansion
}}}{\longleftrightarrow}\,  &  f(x^{4},x^{3},x^{2},x^{1})\,\underline
{\widetilde{\odot}}_{\bar{L}}\,g(y^{4},y^{3},y^{2},y^{1}),\nonumber
\end{align}
symbolizing a transition via the substitutions
\begin{equation}
D_{q^{a}}^{i}\rightarrow D_{q^{-a}}^{i^{\prime}},\quad\hat{n}_{i}%
\rightarrow-\hat{n}_{i^{\prime}},\quad x^{i}\rightarrow x^{i^{\prime}},\quad
q^{\pm1}\rightarrow q^{\mp1},
\end{equation}
where $i^{\prime}\equiv5-i.$

\section{q-Deformed Minkowski space\label{MinSpa}}

In this section we would like to deal with q-deformed Minkowski space
\cite{CSSW90, SWZ91, Maj91, LWW97}, which from a physical point of view is the
most interesting one (for other versions of deformed spacetime see also Refs.
\cite{Lu92, Cas93, Dob94, ChDe95, ChKu04, Koch04}). First of all, let us make
contact with its defining relations. In terms of the coordinates $X^{\mu},$
$\mu\in\{0,+,-,3/0\},$ they take the form%
\begin{align}
X^{\mu}X^{0}  &  =X^{0}X^{\mu},\quad\mu\in\{0,+,-,3/0\},\label{Minkowskialg}\\
X^{\pm}X^{3/0}  &  =q^{\mp2}X^{3/0}X^{\pm},\nonumber\\
X^{-}X^{+}-X^{+}X^{-}  &  =\lambda(X^{3/0}X^{3/0}+X^{0}X^{3/0}).\nonumber
\end{align}
Notice that $X^{0}$ can be interpreted as a time coordinate, while the
$X^{\mu}$ with $\mu=\{+,3/0,-\}$ give some sort of light cone coordinates.

As in the previous sections we start our further considerations with the
braiding of the Minkowski space coordinates \cite{OSWZ92}:
\begin{align}
\Psi_{X,Y}^{-1}(X^{3/0}\otimes w)  &  =(\Lambda^{-1/2}\tau^{1}\triangleright
w)\otimes X^{3/0}\\
&  -\,q^{1/2}\lambda_{+}^{1/2}\lambda(\Lambda^{-1/2}(\tau^{3})^{-1/2}%
S^{1}\triangleright w)\otimes X^{+},\nonumber\\
\Psi_{X,Y}^{-1}(X^{+}\otimes w)  &  =(\Lambda^{-1/2}(\tau^{3})^{-1/2}%
\sigma^{2}\triangleright w)\otimes X^{+}\nonumber\\
&  -q^{\frac{3}{2}}\lambda_{+}^{-1/2}\lambda(\Lambda^{-1/2}T^{2}\triangleright
w)\otimes X^{3/0},\\
\Psi_{X,Y}^{-1}(X^{-}\otimes w)  &  =(\Lambda^{-1/2}(\tau^{3})^{1/2}\tau
^{1}\triangleright w)\otimes X^{-}\nonumber\\
&  -q^{-1/2}\lambda_{+}^{1/2}\lambda(\Lambda^{-1/2}S^{1}\triangleright
w)\otimes X^{0}\\
&  -\,\lambda^{2}(\Lambda^{-1/2}(\tau^{3})^{-1/2}T^{-}S^{1}\triangleright
w)\otimes X^{+}\nonumber\\
&  +\,q^{-1/2}\lambda_{+}^{-1/2}\lambda(\Lambda^{-1/2}(\tau^{1}T^{-}%
-q^{-1}S^{1})\triangleright w)\otimes X^{3/0},\nonumber\\
\Psi_{X,Y}^{-1}(X^{0}\otimes w)  &  =(\Lambda^{-1/2}\sigma^{2}\triangleright
w)\otimes X^{0}\nonumber\\
&  -q^{1/2}\lambda_{+}^{-1/2}\lambda(\Lambda^{-1/2}T^{2}(\tau^{3}%
)^{1/2}\triangleright w)\otimes X^{-}\\
&  +\,q^{1/2}\lambda_{+}^{-1/2}\lambda(\Lambda^{-1/2}(\tau^{3})^{-1/2}%
(T^{-}\sigma^{2}+qS^{1})\triangleright w)\otimes X^{+}\nonumber\\
&  -\,\lambda_{+}^{-1}(\Lambda^{-1/2}(\lambda^{2}T^{-}T^{2}+q(\tau^{1}%
-\sigma^{2}))\triangleright w)\otimes X^{3/0},\nonumber
\end{align}
where $T^{-},$ $T^{2},$ $S^{1},$ $\sigma^{2},$ $\tau^{1},$ and $\tau^{3}$ are
generators of the q-deformed Lorentz algebra (for its algebraic structure see
Appendix \ref{QuanAlg}).

Applying considerations very similar to those for the Euclidean cases we can
derive the identities%
\begin{align}
&  \Psi_{X,Y}^{-1}((X^{+})^{n}\otimes w)=\sum_{k=0}^{n}%
%TCIMACRO{\QATOPD{[}{]}{n}{k}}%
%BeginExpansion
\genfrac{[}{]}{0pt}{}{n}{k}%
%EndExpansion
_{q^{2}}(-q^{3/2}\lambda_{+}^{-1/2}\lambda)^{k}\\
&  \times\,\Big [(\Lambda^{-1/2})^{n}(T^{2})^{k}(\sigma^{2})^{n-\,k}(\tau
^{3})^{-(n-\,k)/2}\triangleright w\Big ]\nonumber\\
&  \otimes\,(X^{+})^{n-k}(X^{3/0})^{k},\nonumber\\[0.16in]
&  \Psi_{X,Y}^{-1}((X^{3/0})^{n}\otimes w)=\sum_{k=0}^{n}%
%TCIMACRO{\QATOPD{[}{]}{n}{k}}%
%BeginExpansion
\genfrac{[}{]}{0pt}{}{n}{k}%
%EndExpansion
_{q^{2}}(-q^{1/2}\lambda_{+}^{1/2}\lambda)^{k}q^{-k(k+1)}\\
&  \times\,\Big [(\Lambda^{-1/2})^{n}(S^{1})^{k}(\tau^{1})^{n-k}(\tau
^{3})^{-k/2}\triangleright w\Big ]\nonumber\\
&  \otimes\,(X^{+})^{k}(X^{3/0})^{n-k}.\nonumber
\end{align}
To get a corresponding expression for powers of $X^{-}$ we have to exploit the
relations%
\begin{align}
&  \Delta_{L}\left(  (T^{-})^{n}\triangleright(X^{3/0})^{n}\right) \\
&  =\sum_{k=0}^{n}%
%TCIMACRO{\QATOPD{[}{]}{n}{k}}%
%BeginExpansion
\genfrac{[}{]}{0pt}{}{n}{k}%
%EndExpansion
_{q^{2}}(-1)^{k}q^{k(k-1)}\,\Delta((T^{-})^{n-k})\,\Delta_{L}((X^{3/0}%
)^{n})\,\Delta((T^{-})^{k}),\nonumber
\end{align}
and
\begin{equation}
\Delta_{L}\left(  (T^{-})^{n}\triangleright(X^{3/0})^{n}\right)
=(q^{3/2}\lambda_{+}^{1/2})^{n}[[n]]_{q^{2}}!\,\Delta_{L}(X^{-})^{n},
\end{equation}
leading us to
\begin{align}
&  \Psi_{X,Y}^{-1}((X^{-})^{n}\otimes w)=\\
&  =\sum_{i=0}^{n}\sum_{k=0}^{n}\sum_{l=0}^{n-k}(-1)^{k+i}\lambda^{i}%
\,\frac{(q^{1/2}\lambda_{+}^{1/2})^{i-n}}{[[n]]_{q^{2}}!}%
\,q^{k(k+1)+i(i-1)+2(n-k-l-\,i)(i+k)-n}\nonumber\\
&  \times\,%
%TCIMACRO{\QATOPD{[}{]}{n}{i}}%
%BeginExpansion
\genfrac{[}{]}{0pt}{}{n}{i}%
%EndExpansion
_{q^{2}}%
%TCIMACRO{\QATOPD{[}{]}{n}{k+l}}%
%BeginExpansion
\genfrac{[}{]}{0pt}{}{n}{k+l}%
%EndExpansion
_{q^{2}}%
%TCIMACRO{\QATOPD{[}{]}{k+l}{k}}%
%BeginExpansion
\genfrac{[}{]}{0pt}{}{k+l}{k}%
%EndExpansion
_{q^{2}}\nonumber\\
&  \times\,\Big [(\Lambda^{-1/2})^{n}(T^{-})^{l}(\tau^{1})^{n-i}(S^{1}%
)^{i}(T^{-})^{k}(\tau^{3})^{(n-\,l-\,i-k)/2}\triangleright w\Big ]\nonumber\\
&  \otimes\,(T^{-})^{n-k-l}\triangleright(X^{+})^{i}(X^{3/0})^{n-i}.\nonumber
\end{align}

There remains to deal with the coordinate $X^{0}$ in the same manner.
Unfortunately, if we try to do so we would get expressions that reveal a
rather complicated structure. Hence, we wish to proceed differently. Towards
this end, we have to recall that in Ref. \cite{Wac04} it was proven that the
monomials
\begin{equation}
(\hat{r}^{2})^{n_{r}}(X^{+})^{n_{+}}(X^{3/0})^{n_{3/0}-\,n_{r}}(X^{-})^{n_{-}%
},\quad n_{\mu}\in\mathbb{N}_{0}, \label{MinBasRa}%
\end{equation}
also constitute a basis of q-deformed Minkowski space, where the new central
coordinate $\hat{r}^{2}$ stands for the square of the Minkowski length given
by%
\begin{equation}
\hat{r}^{2}=-X^{0}X^{0}+X^{3}X^{3}-qX^{+}X^{-}-q^{-1}X^{-}X^{+}.
\end{equation}
Let us notice that in Ref. \cite{Wac04} one can also find the relationship
between this new basis and a more familiar one with $\hat{r}^{2}$ being
replaced by the time coordinate $X^{0}$. The reason for introducing this
special basis lies in the fact that the Minkowski length has a rather simple
braiding, since it satisfies%
\begin{equation}
\Psi_{X,Y}^{-1}((\hat{r}^{2})^{n}\otimes w)=(\Lambda^{-n}\triangleright
w)\otimes\hat{r}^{2n},
\end{equation}
which follows from the very definition of $\hat{r}^{2}.$ Putting everything
together we get in a straightforward manner%
\begin{align}
&  \Psi_{X,Y}^{-1}((\hat{r}^{2})^{n_{r}}(X^{+})^{n_{+}}(X^{3/0})^{n_{3/0}%
}(X^{-})^{n_{-}}\otimes w)=\label{MinBraiMon}\\
&  \frac{(q^{3/2}\lambda_{+}^{1/2})^{-n_{-}}}{[[n_{-}]]_{q^{2}}!}\sum
_{i=0}^{n_{+}}\sum_{j=0}^{n_{3/0}}(-q^{3/2}\lambda_{+}^{1/2}\lambda
)^{i}(-q^{1/2}\lambda_{+}^{1/2}\lambda)^{j}\,q^{-j(j+1)-2j(n_{+}%
-\,2i)}\nonumber\\
&  \times%
%TCIMACRO{\QATOPD{[}{]}{n_{+}}{i}}%
%BeginExpansion
\genfrac{[}{]}{0pt}{}{n_{+}}{i}%
%EndExpansion
_{q^{2}}%
%TCIMACRO{\QATOPD{[}{]}{n_{3/0}}{j}}%
%BeginExpansion
\genfrac{[}{]}{0pt}{}{n_{3/0}}{j}%
%EndExpansion
_{q^{2}}\sum_{l=0}^{n_{-}}\sum_{v=0}^{n_{-}-\,l}(-1)^{l}q^{l(l-1)}%
%TCIMACRO{\QATOPD{[}{]}{l+v}{v}}%
%BeginExpansion
\genfrac{[}{]}{0pt}{}{l+v}{v}%
%EndExpansion
_{q^{2}}%
%TCIMACRO{\QATOPD{[}{]}{n_{-}}{l+v}}%
%BeginExpansion
\genfrac{[}{]}{0pt}{}{n_{-}}{l+v}%
%EndExpansion
_{q^{2}}\nonumber\\
&  \times\sum_{t=0}^{n_{-}}(-q^{1/2}\lambda_{+}^{1/2}\lambda)^{t}q^{-t(t+1)}%
%TCIMACRO{\QATOPD{[}{]}{n_{-}}{t}}%
%BeginExpansion
\genfrac{[}{]}{0pt}{}{n_{-}}{t}%
%EndExpansion
_{q^{2}}\nonumber\\
&  \times\Big[(T^{2})^{i}(S^{1})^{j}(\sigma^{2})^{n_{+}-\,i}(\tau
^{1})^{n_{3/0}-j}(\tau^{3})^{-(n_{+}-\,n_{-}-\,i+j+l+v)/2}\nonumber\\
&  \times(S^{1})^{t}(\tau^{1})^{n_{-}-\,t}(\tau^{3})^{-t/2}(T^{-})^{l}%
(\Lambda^{-1/2})^{2n_{r}+\,n_{+}+\,n_{3/0}+\,n_{-}}\Big]\triangleright
w\nonumber\\
&  \otimes(\hat{r}^{2})^{n_{r}}(X^{+})^{n_{+}-\,i+j}(X^{3/0})^{n_{3/0}%
-j+i}(T^{-})^{n_{-}-\,l-v}\triangleright\left(  (X^{+})^{t}(X^{3/0}%
)^{n_{-}-\,t}\right)  .\nonumber
\end{align}

Our next job is to rearrange the symmetry generators within the square bracket
in a way that generators of the same type are brought together. With the help
of the commutation relations in Appendix \ref{QuanAlg} [see Eqs.
(\ref{LorAlgAnf})-(\ref{LorAlgEnd})] this task can be done rather easily in
the case of the symmetry generator $S^{1}$. For computational convenience it
is also necessary to commute powers of $\sigma^{2}$ and $\tau^{2}$ to the far
left of the product of symmetry generators. To achieve this, we apply the two
identities%
\begin{align}
(\tau^{1})^{n}(T^{-})^{m}  &  =\sum_{k=0}^{\min(n,m)}(-\lambda)^{k}%
\,q^{-2m(n-k)}\,[[k]]_{q^{-2}}!%
%TCIMACRO{\QATOPD{[}{]}{n}{k}}%
%BeginExpansion
\genfrac{[}{]}{0pt}{}{n}{k}%
%EndExpansion
_{q^{2}}%
%TCIMACRO{\QATOPD{[}{]}{m}{k}}%
%BeginExpansion
\genfrac{[}{]}{0pt}{}{m}{k}%
%EndExpansion
_{q^{-2}}\\
&  \times(S^{1})^{k}(T^{-})^{m-k}(\tau^{1})^{n-k},\nonumber\\
(\sigma^{2})^{n}(T^{-})^{m}  &  =\sum_{k=0}^{\min(n,m)}(q^{2}\lambda
)^{k}\,q^{2m(n-k)}\,[[k]]_{q^{2}}!%
%TCIMACRO{\QATOPD{[}{]}{n}{k}}%
%BeginExpansion
\genfrac{[}{]}{0pt}{}{n}{k}%
%EndExpansion
_{q^{-2}}%
%TCIMACRO{\QATOPD{[}{]}{m}{k}}%
%BeginExpansion
\genfrac{[}{]}{0pt}{}{m}{k}%
%EndExpansion
_{q^{2}}\\
&  \times(S^{1})^{k}(T^{-})^{m-k}(\sigma^{2})^{n-k},\nonumber
\end{align}
which can be proven by the method of induction in combination with the
relations of (\ref{LorAlgAnf}) and (\ref{LorAlg2}) in Appendix \ref{QuanAlg}.
With these considerations it follows that%
\begin{align}
&  (T^{2})^{i}(S^{1})^{j}(\sigma^{2})^{n_{+}-\,i}(\tau^{1})^{n_{3/0}-\,j}%
(\tau^{3})^{-(n_{+}-\,n_{-}-\,i+j+l+v)/2}\label{MinProSymGen}\\
&  \times(S^{1})^{t}(\tau^{1})^{n_{-}-\,\,t}(\tau^{3})^{-t/2}(T^{-}%
)^{l}(\Lambda^{-1/2})^{2n_{r}+n_{+}+\,n_{3/0}+\,n_{-}}\nonumber\\
=  &  \,q^{-2l(\,n_{3/0}+t)-2t(n_{+}-\,n_{-}+v-i+j)}\nonumber\\
&  \times\sum_{u_{1}=0}^{l}\sum_{u_{2}=0}^{l-u_{1}}\sum_{u_{3}=0}%
^{l-u_{1}-u_{2}}(-\lambda)^{u_{1}+u_{2}}\lambda^{u_{3}}\,\nonumber\\
&  \times\,q^{-2u_{1}u_{2}+u_{3}(u_{3}+1)+2l(u_{1}+u_{2})}\,q^{-2(n_{+}%
-\,i)(u_{1}+u_{2}+u_{3})+2(n_{3/0}-\,j)u_{1}}\nonumber\\
&  \times\lbrack\lbrack u_{1}+u_{2}+u_{3}]]_{q^{-2}}!%
%TCIMACRO{\QATOPD{[}{]}{l}{u_{1}+u_{2}+u_{3}}}%
%BeginExpansion
\genfrac{[}{]}{0pt}{}{l}{u_{1}+u_{2}+u_{3}}%
%EndExpansion
_{q^{-2}}\nonumber\\
&  \times%
%TCIMACRO{\QATOPD{[}{]}{n_{-}-t}{u_{1}}}%
%BeginExpansion
\genfrac{[}{]}{0pt}{}{n_{-}-t}{u_{1}}%
%EndExpansion
_{q^{2}}%
%TCIMACRO{\QATOPD{[}{]}{n_{3/0}-j}{u_{2}}}%
%BeginExpansion
\genfrac{[}{]}{0pt}{}{n_{3/0}-j}{u_{2}}%
%EndExpansion
_{q^{2}}%
%TCIMACRO{\QATOPD{[}{]}{n_{+}-i}{u_{3}}}%
%BeginExpansion
\genfrac{[}{]}{0pt}{}{n_{+}-i}{u_{3}}%
%EndExpansion
_{q^{2}}\nonumber\\
&  \times(T^{2})^{i}(S^{1})^{j+t+u_{1}+u_{2}+u_{3}}(T^{-})^{l-u_{1}%
-u_{2}-u_{3}}(\sigma^{2})^{n_{+}-\,i-u_{3}}\nonumber\\
&  \times(\tau^{1})^{n_{-}+\,n_{3/0}-\,j-t-u_{1}-u_{2}}(\tau^{3}%
)^{1/2(n_{-}-\,n_{+}+\,i-j-l-v-t)}(\Lambda^{-1/2})^{2n_{r}+n_{+}%
+\,n_{3/0}+\,n_{-}}.\nonumber
\end{align}

A short glance at Eq. (\ref{MinBraiMon}) makes it obvious that we cannot write
down a closed formula for braided products on a commutative algebra until we
have not evaluated the action of powers of $T^{-}$ as it appears in the second
tensor factor of the last expression of (\ref{MinBraiMon}). Using the results
of Appendix \ref{ActSymAlg} [see Eq. (\ref{WirkT-Min})], one can show by
direct calculation that
\begin{align}
&  (\hat{r}^{2})^{n_{r}}(X^{+})^{n_{+}-\,i+j}(X^{3/0})^{n_{3/0}-\,j+i}%
(T^{-})^{n_{-}-\,l-v}\triangleright\left(  (X^{+})^{t}(X^{3/0})^{n_{-}%
-\,t}\right) \label{MinBraiSecTens}\\
&  =\sum_{w=0}^{n_{-}-\,k-l}\sum_{%
%TCIMACRO{\QATOP{0\leq w_{1}+w_{2}\leq\min(w,t)}{w\leq w_{1}+2w_{2}}}%
%BeginExpansion
\genfrac{}{}{0pt}{}{0\leq w_{1}+w_{2}\leq\min(w,t)}{w\leq w_{1}+2w_{2}}%
%EndExpansion
}(q^{3/2}\lambda_{+}^{1/2})^{n_{-}-\,k-l-w}\,q^{-2n_{-}t}\,(d_{q}%
)_{w_{1},w_{2}}^{w,t}\nonumber\\
&  \times q^{2t(k+l+w)+2(w-w_{1}-w_{2})(k+l+w-t)}\,[[n_{-}-k-l-w]]_{q^{2}%
}!\nonumber\\
&  \times%
%TCIMACRO{\QATOPD{[}{]}{n_{-}-t}{n_{-}-k-l-w}}%
%BeginExpansion
\genfrac{[}{]}{0pt}{}{n_{-}-t}{n_{-}-k-l-w}%
%EndExpansion
_{q^{2}}%
%TCIMACRO{\QATOPD{[}{]}{n_{-}-k-l}{w}}%
%BeginExpansion
\genfrac{[}{]}{0pt}{}{n_{-}-k-l}{w}%
%EndExpansion
_{q^{2}}\nonumber\\
&  \times\sum_{p=0}^{w_{1}}q^{-p+2p(k+l-t+w_{1}+2w_{2})}\,q^{2(n_{3/0}%
-\,j+i)(t+p-w_{1}-w_{2})}\nonumber\\
&  \times(\hat{r}^{2})^{n_{r}}(X^{+})^{n_{+}-\,i+j+t-w_{1}-w_{2}+p}%
(X^{3/0})^{n_{3/0}-\,j+\,i+k+l-t+w_{1}+2w_{2}-p}\nonumber\\
&  \times S_{w_{1},p}^{0}(\hat{r}^{2},X^{3/0})(X^{-})^{n_{-}-\,k-l-w_{1}%
-w_{2}+p},\nonumber
\end{align}
where the explicit forms of $(d_{q})_{k,l}^{i,j}$ and $S_{i,j}^{0}$ have been
introduced in Appendix \ref{ActSymAlg} [see Eqs. (\ref{dq}) and (\ref{S0kp})].
As a next step we insert Eqs. (\ref{MinProSymGen}) and (\ref{MinBraiSecTens})
into Eq. (\ref{MinBraiMon}) and evaluate the actions for $\tau^{1},$
$\sigma^{2}$, and $\tau^{3}$ which can easily be read off from the relations
in (\ref{WirkTau}) and (\ref{WirkSigma}) in Appendix \ref{ActSymAlg} [the
action of the scaling operator $\Lambda$ is very simple, as one can see from
Eq. (\ref{LamMin}) in Appendix \ref{QuanAlg}]. Proceeding this way provides us
with an expression we can directly translate into a formula for the braided
product of two commutative functions. If we choose for the algebra isomorphism
of Eq. (\ref{AlgIso})%
\begin{align}
&  \mathcal{W}\left(  (r^{2})^{n_{r}}(x^{+})^{n_{+}}(x^{3/0})^{n_{3/0}}%
(x^{-})^{n_{-}}\right) \\
&  \equiv(\hat{r}^{2})^{n_{r}}(X^{+})^{n_{+}}(X^{3/0})^{n_{3/0}}(X^{-}%
)^{n_{-}},\quad n_{\mu}\in\mathbb{N},\nonumber
\end{align}
we obtain for the braided product after some tedious steps
\begin{align}
&  f(r^{2},x^{+},x^{3/0},x^{-})\,\underline{\odot}_{L}\,g(r^{2},x^{+}%
,x^{3/0},x^{-})\label{MinComBrai}\\
&  =\sum_{i=0}^{\infty}\sum_{j=0}^{\infty}(-q^{3/2}\lambda_{+}^{1/2}%
\lambda)^{i}\,(-q^{1/2}\lambda_{+}^{1/2}\lambda)^{j}\,\frac{q^{-j(j+1)+4ij}%
}{[[i]]_{q^{2}}!\,[[j]]_{q^{2}}!}\nonumber\\
&  \times\sum_{k=0}^{\infty}\sum_{l=0}^{\infty}\sum_{s=0}^{\infty}%
(-1)^{k}\,(q^{-3/2}\lambda_{+}^{-1})^{k+l}\,(-q^{1/2}\lambda_{+}^{1/2}%
\lambda)^{s}\nonumber\\
&  \qquad\times\,q^{k(k-1)-s(s-1)}\,\frac{q^{-2ks+2(i-j)(k+2s)}}{[[l]]_{q^{2}%
}!\,[[s]]_{q^{2}}!}\nonumber\\
&  \times\sum_{t_{1}=0}^{k}\sum_{t_{2}=0}^{k-t_{1}}\sum_{t_{3}=0}%
^{k-t_{1}-t_{2}}(-\lambda)^{t_{1}+t_{2}}(q^{2}\lambda)^{t_{3}}q^{-t_{3}%
(t_{3}-1)}\nonumber\\
&  \qquad\times\,\frac{q^{2(i-k+1)(t_{1}+t_{2}+t_{3})-2k(i-s-t_{1}%
)+2(k-t_{1})(j+t_{2})}}{[[t_{1}]]_{q^{2}}!\,[[t_{2}]]_{q^{2}}!\,[[t_{3}%
]]_{q^{2}}!\,[[k-t_{1}-t_{2}-t_{3}]]_{q^{2}}!}\nonumber\\
&  \times\sum_{u_{1}=0}^{\infty}\sum_{u_{2}=0}^{\infty}q^{-u_{1}%
(u_{1}-1)-u_{2}(u_{2}-1)-4u_{1}u_{2}}\,\frac{q^{-2u_{1}(j+t_{2})-2(u_{1}%
+u_{2})(i+t_{3})}}{[[u_{1}]]_{q^{2}}!\,[[u_{2}]]_{q^{2}}!}\nonumber\\
&  \times\sum_{v=0}^{\infty}\sum_{%
%TCIMACRO{\QATOP{0\leq w_{1}+w_{2}\leq\min(v,s)}{v\leq w_{1}+2w_{2}}}%
%BeginExpansion
\genfrac{}{}{0pt}{}{0\leq w_{1}+w_{2}\leq\min(v,s)}{v\leq w_{1}+2w_{2}}%
%EndExpansion
}(q^{-3/2}\lambda_{+}^{-1/2})^{v}\,(d_{q})_{w_{1},w_{2}}^{v,s}\nonumber\\
&  \qquad\times\,\frac{q^{2sv+2(v-w_{1}-w_{2})(k+l+v-s)-2(w_{1}+w_{2})(i-j)}%
}{[[v]]_{q^{1}}!\,[[k+l+v-s]]_{q^{2}}!}\nonumber\\
&  \times\sum_{p=0}^{w_{1}}q^{-p+2p(i-j+k+l-s+w_{1}+2w_{2})}\nonumber\\
&  \times\Big[(T^{2})^{i}(S^{1})^{j+s+t_{1}+t_{2}+t_{3}}(T^{-})^{k-t_{1}%
-t_{2}-t_{3}}\big((y^{3/0})^{2}D_{q^{2}}^{+}D_{q^{2}}^{-}\big)^{u_{1}+u_{2}%
}\nonumber\\
&  \qquad\otimes(x^{+})^{\alpha_{+}}(x^{3/0})^{\alpha_{3/0}}\,S_{w_{1}%
,\,p}^{0}(r^{2},x^{3/0})\,(x^{-})^{\alpha_{-}}\nonumber\\
&  \qquad\times(D_{q^{2}}^{+})^{i+t_{3}}(D_{q^{2}}^{3/0})^{j+t_{2}+u_{2}%
}(D_{q^{2}}^{-})^{k+l+v-s}(x^{-})^{t_{1}+u_{1}}(D_{q^{2}}^{-})^{s+t_{1}+u_{1}%
}\Big ]\nonumber\\
&  \times q^{(\hat{n}_{+}+\,\hat{n}_{3/0}+\,\hat{n}_{-}+\,2\hat{n}_{r}%
)\otimes(\hat{n}_{+}+\,\hat{n}_{3/0}+\,\hat{n}_{-}+\,2\hat{n}_{r})+2(\hat
{n}_{-}-\,\hat{n}_{+})\otimes(\hat{n}_{-}-\,\hat{n}_{+})}\nonumber\\
&  \times\Big[g(q^{\beta_{+}}y^{+},q^{\beta_{3/0}}y^{3/0},q^{\beta_{-}}%
y^{-})\otimes f(q^{\gamma_{+}}x^{+},q^{\gamma_{3/0}}x^{3/0})\Big ],\nonumber
\end{align}
where, for brevity, we have introduced
\begin{align}
\alpha_{+}  &  \equiv j+s+t_{3}-w_{1}-w_{2}+p,\\
\alpha_{3/0}  &  \equiv i+k+l-s+t_{2}+u_{2}+w_{1}+2w_{2}-p,\nonumber\\
\alpha_{-}  &  \equiv v-w_{1}-w_{2}+p,\nonumber\\[0.16in]
\beta_{+}  &  \equiv-i+j+2(k+l)+s-t_{1}-t_{2}+t_{3},\\
\beta_{3/0}  &  \equiv i-j-s-t_{1}-t_{2}+t_{3}-2(u_{1}+u_{2}),\nonumber\\
\beta_{-}  &  =-\beta_{+},\nonumber\\[0.16in]
\gamma_{+}  &  \equiv-2(s+t_{1}+t_{2}+t_{3}-u_{1}-u_{2}),\\
\gamma_{3/0}  &  \equiv-2(k-s-t_{1}-u_{1}+w_{1}+w_{2}-p).\nonumber
\end{align}
Let us mention that in Eq. (\ref{MinComBrai}) the tensor product of operators
has to act componentwise on the tensor product of commutative functions.
Again, the explicit form of the required actions can be looked up in Appendix
\ref{ActSymAlg} [see Eqs. (\ref{T-fMin})-(\ref{WirkSigma})].

For the sake of completeness, we wish to write down the rule making a
connection to the second braiding which is determined by the relations%
\begin{align}
\Psi_{X,Y}(X^{3/0}\otimes w)  &  =(\Lambda^{1/2}(\tau^{3})^{-1/2}\sigma
^{2}\triangleright w)\otimes X^{3/0}\\
&  -\,q^{\frac{3}{2}}\lambda_{+}^{1/2}\lambda(\Lambda^{1/2}T^{2}\triangleright
w)\otimes X^{-},\nonumber\\
\Psi_{X,Y}(X^{-}\otimes w)  &  =(\Lambda^{1/2}\tau^{1}\triangleright w)\otimes
X^{-}\nonumber\\
&  -q^{1/2}\lambda_{+}^{-1/2}\lambda(\Lambda^{1/2}(\tau^{3})^{-1/2}%
S^{1}\triangleright w)\otimes X^{3/0},\\
\Psi_{X,Y}(X^{+}\otimes w)  &  =(\Lambda^{1/2}\sigma^{2}\triangleright
w)\otimes X^{+}\nonumber\\
&  -q^{1/2}\lambda_{+}^{1/2}\lambda(\Lambda^{1/2}T^{2}(\tau^{3})^{1/2}%
\triangleright w)\otimes X^{0}\\
&  -\,q^{1/2}\lambda_{+}^{-1/2}\lambda(\Lambda^{1/2}(\tau^{3})^{-1/2}%
(T^{+}\sigma^{2}+q\tau^{3}T^{2})\triangleright w)\otimes X^{3/0}\nonumber\\
&  +\,q^{2}\lambda^{2}(\Lambda^{1/2}T^{2}T^{+}\triangleright w)\otimes
X^{-},\nonumber\\
\Psi_{X,Y}(X^{0}\otimes w)  &  =(\Lambda^{1/2}(\tau^{3})^{1/2}\tau
^{1}\triangleright w)\otimes X^{0}\nonumber\\
&  -q^{-1/2}\lambda_{+}^{-1/2}\lambda(\Lambda^{1/2}S^{1}\triangleright
w)\otimes X^{+}\\
&  -\,q^{1/2}\lambda_{+}^{-1/2}\lambda(\Lambda^{1/2}(qT^{+}\tau^{1}%
-T^{2}))\triangleright w)\otimes X^{-}\nonumber\\
&  +\,\lambda_{+}^{-1}(\Lambda^{1/2}(\tau^{3})^{-1/2}(\lambda^{2}T^{+}%
S^{1}+q^{-1}(\tau^{3}\tau^{1}-\sigma^{2}))\triangleright w)\otimes
X^{3/0}.\nonumber
\end{align}
In very much the same way as for the Euclidean spaces we can write%
\begin{align}
&  f(r^{2},x^{+},x^{3/0},x^{-})\,\underline{\odot}_{L}\,g(r^{2},y^{+}%
,y^{3/0},y^{-})\\
\overset{{{%
%TCIMACRO{\QATOP{\pm}{q}}%
%BeginExpansion
\genfrac{}{}{0pt}{}{\pm}{q}%
%EndExpansion
}{%
%TCIMACRO{\QATOP{\rightarrow}{\rightarrow}}%
%BeginExpansion
\genfrac{}{}{0pt}{}{\rightarrow}{\rightarrow}%
%EndExpansion
}{%
%TCIMACRO{\QATOP{\mp}{1/q}}%
%BeginExpansion
\genfrac{}{}{0pt}{}{\mp}{1/q}%
%EndExpansion
}}}{\longleftrightarrow}\,  &  f(x^{-},x^{3/0},x^{+},r^{2})\,\underline
{\widetilde{\odot}}_{\bar{L}}\,g(y^{-},y^{3/0},y^{+},r^{2}),\nonumber
\end{align}
where the transition symbol has the same meaning as in Sec. \ref{Kap2}.

\section{Conclusion\label{Conclusion}}

Let us end with a few comments on our results. In our previous work we started
a program to provide us with explicit formulae for elements of q-deformed
analysis that can be important in formulating physical theories on quantum
spaces. By calculating formulae for braided products we have now finished this
program. In this manner we have constructed nothing else than multidimensional
extensions of the well-known q-calculus (see for example the presentation in
Ref. \cite{Kac00}). In our future work we will show how to use this new
formalism in formulating physical theories. One should also notice that the
expressions we have obtained so far can serve as starting point for
implementing q-analysis on a computer algebra system. This way we will lay the
foundations\ for evaluating physical theories based on quantum symmetries.

However, when we apply the results of this article, we have to take into
account the following observations. First of all, let us note that the
formulae in this article always refer to the braiding of two coordinate
spaces, but they are also valid, if we consider the braiding of two momentum
spaces, since momentum generators show the same algebraic properties as
coordinate generators. Things become slightly different, if we take our
attention to the braiding between a momentum space and a coordinate space. The
reason for this lies in the fact that the scaling operators $\Lambda$ act on
coordinate and momentum generators differently [see Eqs. (\ref{Lam2dim}),
(\ref{Lam3dim}), (\ref{Lam4dim}), and (\ref{LamMin}) in\ Appendix
\ref{QuanAlg}]. Due to this fact our formulae have to be adapted in the
following way:

\begin{enumerate}
\item[a)] (quantum plane)\newline%
\begin{align}
f(\underline{x})\,\underline{\widetilde{\odot}}_{\bar{L}}\,g(\underline
{\partial})  &  =\left.  f(\underline{x})\,\underline{\widetilde{\odot}}%
_{\bar{L}}\,g(\underline{y})\right\vert _{y^{A}\rightarrow q^{-3}\partial^{A}%
},\\
f(\underline{\partial})\,\underline{\widetilde{\odot}}_{\bar{L}}%
\,g(\underline{y})  &  =\left.  f(\underline{x})\,\underline{\widetilde{\odot
}}_{\bar{L}}\,g(\underline{y})\right\vert _{x^{A}\rightarrow q^{-3}%
\partial^{A}},\nonumber\\
f(\underline{x})\,\underline{\odot}_{L}\,g(\underline{\partial})  &  =\left.
f(\underline{x})\,\underline{\odot}_{L}\,g(\underline{y})\right\vert
_{y^{A}\rightarrow q^{3}\partial^{A}},\nonumber\\
f(\underline{\partial})\,\underline{\odot}_{L}\,g(\underline{y})  &  =\left.
f(\underline{x})\,\underline{\odot}_{L}\,g(\underline{y})\right\vert
_{x^{A}\rightarrow q^{3}\partial^{A}},\nonumber
\end{align}

\item[b)] (three-dimensional Euclidean space)\newline%
\begin{align}
f(\underline{x})\,\underline{\odot}_{\bar{L}}\,g(\underline{\partial})  &
=\left.  f(\underline{x})\,\underline{\odot}_{\bar{L}}\,g(\underline
{y})\right\vert _{y^{A}\rightarrow q^{-4}\partial^{A}},\\
f(\underline{\partial})\,\underline{\odot}_{\bar{L}}\,g(\underline{y})  &
=\left.  f(\underline{x})\,\underline{\odot}_{\bar{L}}\,g(\underline
{y})\right\vert _{x^{A}\rightarrow q^{-4}\partial^{A}},\nonumber\\
f(\underline{x})\,\underline{\widetilde{\odot}}_{L}\,g(\underline{\partial})
&  =\left.  f(\underline{x})\,\underline{\widetilde{\odot}}_{L}\,g(\underline
{y})\right\vert _{y^{A}\rightarrow q^{4}\partial^{A}},\nonumber\\
f(\underline{\partial})\,\underline{\widetilde{\odot}}_{L}\,g(\underline{y})
&  =\left.  f(\underline{x})\,\underline{\widetilde{\odot}}_{L}\,g(\underline
{y})\right\vert _{x^{A}\rightarrow q^{4}\partial^{A}},\nonumber
\end{align}

\item[c)] (four-dimensional Euclidean space)\newline%
\begin{align}
f(\underline{x})\,\underline{\widetilde{\odot}}_{\bar{L}}\,g(\underline
{\partial})  &  =\left.  f(\underline{x})\,\underline{\widetilde{\odot}}%
_{\bar{L}}\,g(\underline{y})\right\vert _{y^{A}\rightarrow q^{-2}\partial^{A}%
},\\
f(\underline{\partial})\,\underline{\widetilde{\odot}}_{\bar{L}}%
\,g(\underline{y})  &  =\left.  f(\underline{x})\,\underline{\widetilde{\odot
}}_{\bar{L}}\,g(\underline{y})\right\vert _{x^{A}\rightarrow q^{-2}%
\partial^{A}},\nonumber\\
f(\underline{x})\,\underline{\odot}_{L}\,g(\underline{\partial})  &  =\left.
\,f(\underline{x})\,\underline{\odot}_{L}\,g(\underline{y})\right\vert
_{y^{A}\rightarrow q^{2}\partial^{A}},\nonumber\\
f(\underline{\partial})\,\underline{\odot}_{L}\,g(\underline{y})  &  =\left.
\,f(\underline{x})\,\underline{\odot}_{L}\,g(\underline{y})\right\vert
_{x^{A}\rightarrow q^{2}\partial^{A}},\nonumber
\end{align}

\item[d)] (q-deformed Minkowski space)\newline%
\begin{align}
f(\underline{x})\,\underline{\widetilde{\odot}}_{\bar{L}}\,g(\underline
{\partial})  &  =\left.  f(\underline{x})\,\underline{\widetilde{\odot}}%
_{\bar{L}}\,g(\underline{y})\right\vert _{y^{A}\rightarrow q^{2}\partial^{A}%
},\\
f(\underline{\partial})\,\underline{\widetilde{\odot}}_{\bar{L}}%
\,g(\underline{y})  &  =\left.  f(\underline{x})\,\underline{\widetilde{\odot
}}_{\bar{L}}\,g(\underline{y})\right\vert _{x^{A}\rightarrow q^{2}\partial
^{A}},\nonumber\\
f(\underline{x})\,\underline{\odot}_{L}\,g(\underline{\partial})  &  =\left.
\,f(\underline{x})\,\underline{\odot}_{L}\,g(\underline{y})\right\vert
_{y^{A}\rightarrow q^{-2}\partial^{A}},\nonumber\\
f(\underline{\partial})\,\underline{\odot}_{L}\,g(\underline{y})  &  =\left.
\,f(\underline{x})\,\underline{\odot}_{L}\,g(\underline{y})\right\vert
_{x^{A}\rightarrow q^{-2}\partial^{A}}.\nonumber
\end{align}

\end{enumerate}

\noindent\textbf{Acknowledgements}\newline First of all I am very grateful to
Eberhard Zeidler for interesting and useful discussions, his special interest
in my work and his financial support. Also I wish to express my gratitude to
Julius Wess for his efforts, suggestions and discussions. Furthermore, I would
like to thank Alexander Schmidt, Fabian Bachmaier, Michael Wohlgenannt, and
Florian Koch for useful discussions and their steady support. Finally, I thank
Dieter L\"{u}st for kind hospitality.

\appendix

\section{Quantum algebras\label{QuanAlg}}

In this Appendix we list for the quantum algebras we are interested in their
defining relations and their Hopf structure.

We denote by $U_{q}(su_{2})$ the algebra over $\mathbb{C}$ generated by three
generators $L^{+},$ $L^{-},$ and $\tau^{3}$ subject to the relations
\cite{LWW97}
\begin{align}
\tau^{3}L^{\pm}  &  =q^{\mp4}L^{\pm}\tau^{3},\\
qL^{+}L^{-}-q^{-1}L^{-}L^{+}  &  =q\lambda^{-1}\lambda_{+}^{-1}(1-\tau
^{-1}).\nonumber
\end{align}
Sometimes it is more convenient to use the generators
\begin{align}
T^{\pm}  &  =q^{\mp1/2}\lambda_{+}^{1/2}(\tau^{3})^{1/2}L^{\pm},\\
T^{3}  &  =\lambda^{-1}(1-\tau^{3}),\nonumber
\end{align}
which satisfy the relations
\begin{align}
q^{-1}T^{+}T^{-}-qT^{-}T^{+}  &  =T^{3},\\
q^{2}T^{3}T^{+}-q^{-2}T^{+}T^{3}  &  =\lambda_{+}T^{+},\nonumber\\
q^{2}T^{-}T^{3}-q^{-2}T^{3}T^{-}  &  =\lambda_{+}T^{-}.\nonumber
\end{align}
The Hopf structure of the generators $L^{+}$, $L^{-},$ and $\tau^{3}$ is given
by
\begin{align}
\Delta(L^{\pm})  &  =L^{\pm}\otimes(\tau^{3})^{-1/2}+1\otimes L^{\pm},\\
\Delta(\tau^{3})  &  =\tau^{3}\otimes\tau^{3},\nonumber\\[0.16in]
S(L^{\pm})  &  =-L^{\pm}(\tau^{3})^{1/2},\\
S(\tau^{3})  &  =(\tau^{3})^{-1},\nonumber\\[0.16in]
\varepsilon(L^{\pm})  &  =0,\\
\varepsilon(\tau^{3})  &  =1,\nonumber
\end{align}
and likewise for\ $T^{+}$, $T^{-},$ and $T^{3},$%
\begin{align}
\Delta(T^{3})  &  =T^{3}\otimes1+\tau^{3}\otimes T^{3},\\
\Delta(T^{\pm})  &  =T^{\pm}\otimes1+(\tau^{3})^{1/2}\otimes T^{\pm
},\nonumber\\[0.16in]
S(T^{3})  &  =-(\tau^{3})^{-1}T^{3},\\
S(T^{\pm})  &  =-(\tau^{3})^{-1/2}T^{\pm},\nonumber\\[0.16in]
\varepsilon(T^{A})  &  =0,\quad A\in\{+,3,-\}.
\end{align}
For our purposes, it is necessary to extend $U_{q}(su_{2})$ by a grouplike
scaling operator being central in the algebra. With coordinates and partial
derivatives it shows the commutation relations%
\begin{align}
\Lambda X^{\alpha}  &  =q^{-2}X^{\alpha}\Lambda,\quad\alpha
=1,2,\label{Lam2dim}\\
\Lambda\partial^{\,\alpha}  &  =q^{2}\partial^{\,\alpha}\Lambda
,\nonumber\\[0.16in]
\Lambda X^{A}  &  =q^{4}X^{A}\Lambda,\quad A\in\{+,3,-\},\label{Lam3dim}\\
\Lambda\partial^{A}  &  =q^{-4}\partial^{A}\Lambda,\nonumber
\end{align}
where $\alpha$ and $A$ denote spinor and vector indices, respectively.

Next, we come to the quantum algebra $U_{q}(so_{4})$, which from an algebraic
point of view is isomorphic to the tensor product of two $U_{q}(su_{2}%
)$-algebras, i.e.
\begin{equation}
U_{q}(so_{4})\cong U_{q}(su_{2})\otimes U_{q}(su_{2}).
\end{equation}
Thus $U_{q}(so_{4})$ is spanned by two commuting sets of $U_{q}(su_{2}%
)$-generators, denoted by $L_{i}^{\pm},$ $K_{i},$ $i=1,2.$ The commutation
relations between generators of the same lower index explicitly read
\cite{Oca96}
\begin{align}
q^{-1}L_{i}^{+}L_{i}^{-}-qL_{i}^{-}L_{i}^{+}  &  =\lambda^{-1}(1-K_{i}%
^{-1}),\\
L_{i}^{\pm}K_{i}  &  =q^{\mp2}K_{i}L_{i}^{\pm},\qquad i=1,2.\nonumber
\end{align}
The corresponding Hopf structure is given by%
\begin{align}
\Delta(K_{i})  &  =K_{i}\otimes K_{i},\\
\Delta(L_{i}^{\pm})  &  =L_{i}^{\pm}\otimes1+K_{i}^{-1}\otimes L_{i}^{\pm
},\nonumber\\[0.16in]
S(K_{i})  &  =K_{i}^{-1},\\
S(L_{i})  &  =-K_{i}L_{i}^{\pm},\nonumber\\[0.16in]
\varepsilon(K_{i})  &  =1,\\
\varepsilon(L_{i}^{\pm})  &  =0.\nonumber
\end{align}
In this case the central scaling operator $\Lambda$ has to satisfy%
\begin{align}
\Lambda X^{i}  &  =q^{2}X^{i}\Lambda,\quad i=1,\ldots,4,\label{Lam4dim}\\
\Lambda\partial^{\,i}  &  =q^{-2}\partial^{\,i}\Lambda.\nonumber
\end{align}

Finally, we would like to consider a q-deformed version of Lorentz algebra, as
it is presented in Refs. \cite{SWZ91} and \cite{OSWZ92}. Its seven generators
are denoted by $T^{+},$ $T^{-},$ $S^{1},$ $T^{2},$ $\tau^{3},$ $\tau^{1},$ and
$\sigma^{2}$. Let us note that the generators $T^{+},$ $T^{-},$ and $\tau^{3}$
generate an $U_{q}(su_{2})$-subalgebra. In addition to the commutation
relations of the $U_{q}(su_{2})$-subalgebra we have the relations
\begin{align}
\tau^{1}T^{+}  &  =T^{+}\tau^{1}+\lambda T^{2},\label{LorAlgAnf}\\
\tau^{1}T^{-}  &  =q^{-2}T^{-}\tau^{1}-\lambda S^{1},\nonumber\\
\tau^{1}T^{2}  &  =q^{2}T^{2}\tau^{1},\nonumber\\
\tau^{1}S^{1}  &  =S^{1}\tau^{1},\nonumber\\[0.16in]
\sigma^{2}T^{+}  &  =T^{+}\sigma^{2}-q^{2}\lambda\tau^{3}T^{2},
\label{LorAlg2}\\
\sigma^{2}T^{-}  &  =q^{2}T^{-}\sigma^{2}+q^{2}\lambda S^{1},\nonumber\\
\sigma^{2}T^{2}  &  =q^{-2}T^{2}\sigma^{2},\nonumber\\
\sigma^{2}S^{1}  &  =S^{1}\sigma^{2},\nonumber\\[0.16in]
T^{+}T^{2}  &  =q^{-2}T^{2}T^{+},\\
T^{-}T^{2}  &  =T^{2}T^{-}+\lambda^{-1}(\sigma^{2}-\tau^{1}),\nonumber\\
T^{+}S^{1}  &  =q^{2}S^{1}T^{+}+\lambda^{-1}(\tau^{3}\tau^{1}-\sigma
^{2}),\nonumber\\
T^{-}S^{1}  &  =S^{1}T^{-},\nonumber\\
T^{+}T^{-}  &  =q^{2}T^{-}T^{+}+q\lambda^{-1}(1-\tau^{3}),\nonumber\\
T^{2}S^{1}  &  =S^{1}T^{2},\nonumber\\[0.16in]
\tau^{1}\sigma^{2}  &  =\sigma^{2}\tau^{1}+q\lambda^{3}T^{2}S^{1}%
,\label{LorAlgEnd}\\
\tau^{3}\tau^{1}  &  =\tau^{1}\tau^{3},\nonumber\\
\tau^{3}\sigma^{2}  &  =\sigma^{2}\tau^{3},\nonumber\\
\tau^{3}T^{\pm}  &  =q^{\mp4}T^{\pm}\tau^{3},\nonumber\\
\tau^{3}T^{2}  &  =q^{-4}T^{2}\tau^{3},\nonumber\\
\tau^{3}S^{1}  &  =q^{4}S^{1}\tau^{3}.\nonumber
\end{align}
The Hopf structure of $S^{1},$ $T^{2},$ $\tau^{1},$ and $\sigma^{2}$ is given
by
\begin{align}
\Delta(\tau^{1})  &  =\tau^{1}\otimes\tau^{1}+\lambda^{2}S^{1}(\tau
^{3})^{-1/2}\otimes T^{2},\\
\Delta(\sigma^{2})  &  =\sigma^{2}\otimes\sigma^{2}+\lambda^{2}T^{2}(\tau
^{3})^{1/2}\otimes S^{1},\nonumber\\
\Delta(T^{2})  &  =T^{2}\otimes\tau^{1}+(\tau^{3})^{-1/2}\sigma^{2}\otimes
T^{2},\nonumber\\
\Delta(S^{1})  &  =S^{1}\otimes\sigma^{2}+(\tau^{3})^{1/2}\tau^{1}\otimes
S^{1},\nonumber\\[0.16in]
S(T^{2})  &  =-q^{2}(\tau^{3})^{1/2}T^{2},\\
S(S^{1})  &  =-(\tau^{3})^{-1/2}S^{1},\nonumber\\
S(\tau^{1})  &  =\sigma^{2},\nonumber\\
S(\sigma^{2})  &  =\tau^{1},\nonumber\\[0.16in]
\varepsilon(\tau^{1})  &  =\varepsilon(\sigma^{2})=1,\\
\varepsilon(T^{2})  &  =\varepsilon(S^{1})=0.\nonumber
\end{align}
For the scaling operator we now have%
\begin{align}
\Lambda X^{\mu}  &  =q^{-2}X^{\mu}\Lambda,\quad\mu\in
\{+,0,3/0,-\},\label{LamMin}\\
\Lambda\partial^{\,\mu}  &  =q^{2}\partial^{\,\mu}\Lambda.\nonumber
\end{align}

\section{\label{ActSymAlg}Actions of symmetry generators}

In this appendix we present formulae for calculating actions of powers of
symmetry generators on normal ordered monomials. Especially for the
computations in Sec. \ref{Kap2} we need to know the action of powers of the
$U_{q}(su_{2})$-generator $L^{-}.$ With the help of Eq. (\ref{CoProLmin}) we
can write at first%
\begin{align}
&  (L^{-})^{n}\triangleright(X^{+})^{m_{+}}(X^{3})^{m_{3}}(X^{-})^{m_{-}}\\
&  =\big [((L^{-})^{n})_{(1)}\triangleright(X^{+})^{m_{+}}\big ]\big [((L^{-}%
)^{n})_{(2)}\triangleright(X^{3})^{m_{3}}(X^{-})^{m_{-}}\big ]\nonumber\\
&  =\sum_{k=0}^{n}q^{-2k(n-k)}%
%TCIMACRO{\QATOPD{[}{]}{n}{k}}%
%BeginExpansion
\genfrac{[}{]}{0pt}{}{n}{k}%
%EndExpansion
_{q^{2}}\big [(L^{-})^{k}\triangleright(X^{+})^{m_{+}}\big ]\nonumber\\
&  \times\big [(L^{-})^{n-k}(\tau^{3})^{-k/2}\triangleright(X^{3})^{m_{3}%
}(X^{-})^{m_{-}}\big ].\nonumber
\end{align}
The action contained in the second square bracket is a rather simple one as
can be seen from direct inspection of the commutation relations between
$L^{-}$ and the quantum space coordinates (for their explicit form see for
example Ref. \cite{BW01}). In the case of the action in the first square
bracket the same commutation relations show us that we can make as ansatz
\begin{equation}
(L^{-})^{n}\triangleright(X^{+})^{m}=\sum_{{%
%TCIMACRO{\QATOP{i+j=n}{0\leq j\leq i\leq m}}%
%BeginExpansion
\genfrac{}{}{0pt}{}{i+j=n}{0\leq j\leq i\leq m}%
%EndExpansion
}}C\,_{i,j}^{n,m}\,(X^{+})^{m-i}(X^{3})^{i-j}(X^{-})^{j},
\end{equation}
which for the unknown coefficients leads to the recursion relation%
\begin{align}
C\,_{i,j}^{n+1,m}  &  =q^{-2j}\,[[m-i+1]]_{q^{4}}\,C\,_{i-1,j}^{n,m}\\
&  +\,q^{-2(j-1)-1}\,[[i-j+1]]_{q^{2}}\,C\,_{i,j-1}^{n,m}.\nonumber
\end{align}
By the method of inserting one can prove that a solution is given by%
\begin{equation}
C\,_{i,j}^{n,m}=q^{-j^{2}}\,t_{i,j}\,[[i]]_{q^{4}}!%
%TCIMACRO{\QATOPD{[}{]}{m}{i}}%
%BeginExpansion
\genfrac{[}{]}{0pt}{}{m}{i}%
%EndExpansion
_{q^{4}},
\end{equation}
with
\begin{align}
t_{i,0}  &  =1,\label{tuv}\\
t_{i,j}  &  =\sum_{0\leq s_{1}+\ldots+s_{j}\leq i}\,\prod_{l=1}^{j}%
q^{-2(j-l+1)s_{l}}\,[[i-j+2l-s_{1}-\ldots-s_{l}]]_{q^{2}}.\nonumber
\end{align}
Putting everything together we arrive at the expression%
\begin{align}
&  (L^{-})^{n}\triangleright(X^{+})^{m_{+}}(X^{3})^{m_{3}}(X^{-})^{m_{-}%
}\label{WirkLMon3dim}\\
&  =\sum_{{%
%TCIMACRO{\QATOP{i+j+k=n}{0\leq j\leq i\leq m_{+}}}%
%BeginExpansion
\genfrac{}{}{0pt}{}{i+j+k=n}{0\leq j\leq i\leq m_{+}}%
%EndExpansion
}}q^{-2k(n-k+j)-k^{2}-2m_{-}n+2j(m_{3}-i-j)-j^{2}}\,t_{i,j}\nonumber\\
&  \times\,[[i]]_{q^{4}}!\,[[k]]_{q^{2}}!%
%TCIMACRO{\QATOPD{[}{]}{n}{k}}%
%BeginExpansion
\genfrac{[}{]}{0pt}{}{n}{k}%
%EndExpansion
_{q^{2}}%
%TCIMACRO{\QATOPD{[}{]}{m_{+}}{i}}%
%BeginExpansion
\genfrac{[}{]}{0pt}{}{m_{+}}{i}%
%EndExpansion
_{q^{4}}%
%TCIMACRO{\QATOPD{[}{]}{m_{3}}{k}}%
%BeginExpansion
\genfrac{[}{]}{0pt}{}{m_{3}}{k}%
%EndExpansion
_{q^{2}}\nonumber\\
&  \times\,(X^{+})^{m_{+}-\,i}(X^{3})^{m_{3}-\,k+i-j}(X^{-})^{m_{-}%
+\,k+j},\nonumber
\end{align}
which, in turn, gives rise to the following action on commutative functions:%
\begin{align}
&  (L^{-})^{n}\triangleright f(x^{+},x^{3},x^{-})\label{WirL-Func}\\
&  =\sum_{{%
%TCIMACRO{\QATOP{i+j+k=n}{0\leq j\leq i}}%
%BeginExpansion
\genfrac{}{}{0pt}{}{i+j+k=n}{0\leq j\leq i}%
%EndExpansion
}}q^{-k(2n-k+2j)-j^{2}}\,t_{i,j}\,%
%TCIMACRO{\QATOPD{[}{]}{n}{k}}%
%BeginExpansion
\genfrac{[}{]}{0pt}{}{n}{k}%
%EndExpansion
_{q^{2}}\nonumber\\
&  \times\,(x^{3})^{i-j}(x^{-})^{k+j}\,(D_{q^{4}}^{+})^{i}(D_{q^{2}}^{3}%
)^{k}\,f(q^{2j}x^{3},q^{-2n}x^{-}).\nonumber
\end{align}

The calculations in Sec. \ref{Kap3} require explicit formulae for the action
of powers of the $U_{q}(so_{4})$-generators $L_{i}^{+},$ $i=1,2.$ Since all of
the above considerations carry over to that case we limit ourselves to
presenting the results without any proof. Using commutation relations listed
in Ref. \cite{BW01} we found for the action on normal ordered monomials
\begin{align}
&  (L_{1}^{+})^{n}\triangleright(X^{1})^{m_{1}}(X^{2})^{m_{2}}(X^{3})^{m_{3}%
}(X^{4})^{m_{4}}\label{L1+Wirk}\\
&  =\sum_{k=0}^{n}(-1)^{k}q^{m_{1}(n-2k)-(m_{2}+m_{3}%
)(n-k)+k(k-1)/2+(n-k)(n-k-1)/2}\nonumber\\
&  \times\,[[k]]_{q^{2}}!\,[[n-k]]_{q^{2}}!%
%TCIMACRO{\QATOPD{[}{]}{n}{k}}%
%BeginExpansion
\genfrac{[}{]}{0pt}{}{n}{k}%
%EndExpansion
_{q^{-2}}%
%TCIMACRO{\QATOPD{[}{]}{m_{1}}{k}}%
%BeginExpansion
\genfrac{[}{]}{0pt}{}{m_{1}}{k}%
%EndExpansion
_{q^{2}}%
%TCIMACRO{\QATOPD{[}{]}{m_{3}}{n-k}}%
%BeginExpansion
\genfrac{[}{]}{0pt}{}{m_{3}}{n-k}%
%EndExpansion
_{q^{2}}\nonumber\\
&  \times\,(X^{1})^{m_{1}-k}(X^{2})^{m_{2}+k}(X^{3})^{m_{3}-(n-k)}%
(X^{4})^{m_{4}+(n-k)},\nonumber\\[0.16in]
&  (L_{2}^{+})^{n}\triangleright(X^{1})^{m_{1}}(X^{2})^{m_{2}}(X^{3})^{m_{3}%
}(X^{4})^{m_{4}}\label{L2+Wirk}\\
&  =\sum_{k=0}^{n}(-1)^{k}q^{m_{1}(n-2k)-(m_{2}+m_{3}%
)(n-k)+k(k-1)/2+(n-k)(n-k-1)/2}\nonumber\\
&  \times\,[[k]]_{q^{2}}!\,[[n-k]]_{q^{2}}!%
%TCIMACRO{\QATOPD{[}{]}{n}{k}}%
%BeginExpansion
\genfrac{[}{]}{0pt}{}{n}{k}%
%EndExpansion
_{q^{-2}}%
%TCIMACRO{\QATOPD{[}{]}{m_{1}}{k}}%
%BeginExpansion
\genfrac{[}{]}{0pt}{}{m_{1}}{k}%
%EndExpansion
_{q^{2}}%
%TCIMACRO{\QATOPD{[}{]}{m_{2}}{n-k}}%
%BeginExpansion
\genfrac{[}{]}{0pt}{}{m_{2}}{n-k}%
%EndExpansion
_{q^{2}}\nonumber\\
&  \times\,(X^{1})^{m_{1}-k}(X^{2})^{m_{2}-(n-k)}(X^{3})^{m_{3}+k}%
(X^{4})^{m_{4}+(n-k)}.\nonumber
\end{align}
These results imply the following actions on commutative functions:
\begin{align}
&  (L_{1}^{+})^{n}\triangleright f(x^{1},x^{2},x^{3},x^{4})\label{ActfL1+}\\
&  =\sum_{k=0}^{n}(-1)^{n-k}q^{k(k-1)/2+(n-k)(n-k-1)/2}(x^{2})^{n-k}%
(x^{4})^{k}\nonumber\\
&  \times\,(D_{q^{2}}^{1})^{n-k}(D_{q^{2}}^{3})^{k}f(q^{-(n-2k)}x^{1}%
,q^{-k}x^{2},q^{-k}x^{3}),\nonumber\\[0.16in]
&  (L_{2}^{+})^{n}\triangleright f(x^{1},x^{2},x^{3},x^{4})\label{ActfL2+}\\
&  =\sum_{k=0}^{n}(-1)^{n-k}q^{k(k-1)/2+(n-k)(n-k-1)/2}(x^{3})^{n-k}%
(x^{4})^{k}\nonumber\\
&  \times\,(D_{q^{2}}^{1})^{n-k}(D_{q^{2}}^{2})^{k}f(q^{-(n-2k)}x^{1}%
,q^{-k}x^{2},q^{-k}x^{3}).\nonumber
\end{align}

Now, we come to the actions being relevant for q-deformed Minkowski space. In
very much the same way as was done for $L^{-}$ in the case of
three-dimensional q-deformed Euclidean space, one can derive the action of
powers of $T^{-}$ on normal ordered monomials. By making use of the coproduct
of $T^{-}$ we get%
\begin{align}
&  (T^{-})^{m}\triangleright(\hat{r}^{2})^{n_{r}}(X^{+})^{n_{+}}%
(X^{3/0})^{n_{3/0}}(X^{-})^{n_{-}}\label{AllWirkT-}\\
&  =\sum_{k=0}^{m}%
%TCIMACRO{\QATOPD{[}{]}{m}{k}}%
%BeginExpansion
\genfrac{[}{]}{0pt}{}{m}{k}%
%EndExpansion
_{q^{2}}(\hat{r}^{2})^{n_{r}}\big [(T^{-})^{m-k}(\tau^{3})^{k/2}%
\triangleright(X^{+})^{n_{+}}\big ]\nonumber\\
&  \times\,\big [(T^{-})^{k}\triangleright(X^{3/0})^{n_{3/0}}(X^{-})^{n_{-}%
}\big ],\nonumber
\end{align}
where we have taken into account that the Minkowski length commutes with all
Lorentz generators (for the explicit form of commutation relations between
Lorentz generators and Minkowski space coordinates see for example Ref.
\cite{BW01}). The action in\ the second bracket takes on a rather simple form,
since it is given by%
\begin{align}
&  (T^{-})^{k}\triangleright(X^{3/0})^{n_{3/0}}(X^{-})^{n_{-}}%
\label{WirT-X3X-}\\
&  =(q^{3/2}\lambda_{+}^{1/2})^{k}\,[[k]]_{q^{2}}!%
%TCIMACRO{\QATOPD{[}{]}{n_{3/0}}{k}}%
%BeginExpansion
\genfrac{[}{]}{0pt}{}{n_{3/0}}{k}%
%EndExpansion
_{q^{2}}(X^{3/0})^{n_{3/0}-\,k}(X^{-})^{n_{-}+\,k}.\nonumber
\end{align}
For the action in the first bracket we can assume the general form
\begin{align}
&  (T^{-})^{m}\triangleright(X^{+})^{n}\label{ExplWirkT-aufX+}\\
&  =\sum_{%
%TCIMACRO{\QATOP{0\leq s+l\leq\min(n,m)}{m\leq s+2l}}%
%BeginExpansion
\genfrac{}{}{0pt}{}{0\leq s+l\leq\min(n,m)}{m\leq s+2l}%
%EndExpansion
}(d_{q})_{s,l}^{m,n}\,(X^{+})^{n-s-l}(X^{0})^{s}(X^{3/0})^{s+2l-m}%
(X^{-})^{m-s-l}.\nonumber
\end{align}
If we introduce new coefficients $\tilde{d}_{q}$ being subject to
\begin{equation}
(d_{q})_{s,l}^{m,n}=(q\lambda_{+})^{m/2}q^{m-s-l}(\tilde{d}_{q})_{s,l,m-s-l}%
^{m,n}, \label{dq}%
\end{equation}
we find the recursion relations%
\begin{align}
(\tilde{d}_{q})_{s,l,i}^{m+1,n}  &  =[[n-s-l-i+1]]_{q^{-2}}\,(\tilde{d}%
_{q})_{s-1,l,i}^{m,n}\\
&  +\,q^{-2(n-s-l-i)}\,[[n-s-l-i+1]]_{q^{4}}\,(\tilde{d}_{q})_{s,l-1,i}%
^{m,n}\nonumber\\
&  +\,q^{-2(n-s-l-i)}\,[[l+1]]_{q^{2}}\,(\tilde{d}_{q})_{s,l+1,i-1}%
^{m,n},\nonumber\\
(\tilde{d}_{q})_{n,0,0}^{n,n}  &  =[[n]]_{q^{-2}}!,
\end{align}
which are solved by%
\begin{equation}
\hspace*{-0.3in}(\tilde{d}_{q})_{s,l,i}^{m,n}=q^{-2i^{2}}(q\lambda_{+}%
)^{-l}\,(T_{q})_{s,l,i}^{n}\,[[s+l]]_{q^{-2}}!%
%TCIMACRO{\QATOPD{[}{]}{n}{s+l}}%
%BeginExpansion
\genfrac{[}{]}{0pt}{}{n}{s+l}%
%EndExpansion
_{q^{-2}},
\end{equation}
with%
\begin{align}
\hspace*{-0.3in}(T_{q})_{s,l,i}^{n}  &  \equiv\sum_{%
%TCIMACRO{\QATOP{s_{1}+\ldots+s_{i+1}=s}{l_{1}+\ldots+l_{i+1}=l}}%
%BeginExpansion
\genfrac{}{}{0pt}{}{s_{1}+\ldots+s_{i+1}=s}{l_{1}+\ldots+l_{i+1}=l}%
%EndExpansion
}\prod_{k=1}^{i}q^{-2\sum_{j=1}^{k}(s_{j}+l_{j})}\,[[l-i+k-l_{1}-\ldots
-l_{k}]]_{q^{2}}\\
&  \times\sum_{%
%TCIMACRO{\QATOP{0\leq t_{\alpha}\leq l_{\alpha}}{\alpha=1,\ldots,i+1}}%
%BeginExpansion
\genfrac{}{}{0pt}{}{0\leq t_{\alpha}\leq l_{\alpha}}{\alpha=1,\ldots,i+1}%
%EndExpansion
}q^{2\sum_{u=1}^{i+1}t_{u}(n-s-l)}\,\prod_{k=1}^{i+1}\Big (q^{2t_{k}\sum
_{u=1}^{k-1}(s_{u}+t_{u})}\,(P_{q})_{t_{k}}^{s_{k},l_{k}}\Big ),\nonumber\\
\hspace*{-0.3in}(P_{q})_{k}^{v,p}  &  \equiv\sum_{%
%TCIMACRO{\QATOP{0\leq j_{1}+\ldots+j_{p}\leq v}{0\leq t_{1}+\ldots+t_{k}\leq
%p}}%
%BeginExpansion
\genfrac{}{}{0pt}{}{0\leq j_{1}+\ldots+j_{p}\leq v}{0\leq t_{1}+\ldots
+t_{k}\leq p}%
%EndExpansion
}q^{2\sum_{u=1}^{k}(j_{1}+\ldots+j_{t_{u}}+\,t_{u})}. \label{ZusatzOpSchluß}%
\end{align}
There remains to replace the time coordinate $X^{0}$ by the square of the
Minkowski length $\hat{r}^{2}$ in Eq. (\ref{ExplWirkT-aufX+}). In
Ref.\ \cite{Wac04} we already addressed this problem and suggested as solution%
\begin{align}
&  (X^{+})^{n_{+}}(X^{3/0})^{n_{3/0}}(X^{0})^{n_{0}}(X^{-})^{n_{-}%
}\label{TraX0R2}\\
&  =\sum_{p=0}^{n_{0}}q^{(2n_{3/0}-1)p}\,(X^{+})^{n_{+}+\,p}(X^{3/0}%
)^{n_{3/0}-\,p}\,S_{n_{0},\,p}^{0}(\hat{r}^{2},X^{3/0})\,(X^{-})^{n_{-}%
+\,p},\nonumber
\end{align}
where%
\begin{align}
&  S_{k,p}^{0}(\hat{r}^{2},X^{3/0})\label{S0kp}\\
&  =%
\begin{cases}
\sum_{j_{1}=0}^{p}\sum_{j_{2}=0}^{j_{1}}\ldots\sum_{j_{k-p}=0}^{j_{k-p-1}%
}\,a_{0}(\hat{r}^{2},q^{2j_{l}}X^{3/0}), & \text{if }0\leq p<k\text{ },\\
1, & \text{if }p=k\text{,}%
\end{cases}
\nonumber
\end{align}
and
\begin{equation}
a_{0}(\hat{r}^{2},X^{3/0})=-\lambda_{+}^{-1}(q\,\hat{r}^{2}(X^{3/0}%
)^{-1}+q^{-1}X^{3/0}).
\end{equation}
From what we have done so far, it is now rather straightforward to show that%
\begin{align}
&  (T^{-})^{m}\triangleright(\hat{r}^{2})^{n_{r}}(X^{+})^{n_{+}}%
(X^{3/0})^{n_{3/0}}(X^{-})^{n_{-}}\label{WirkT-Min}\\
&  =\sum_{k=0}^{m}\sum_{%
%TCIMACRO{\QATOP{0\leq s+t\leq\min(k,n_{+})}{k\leq s+2t}}%
%BeginExpansion
\genfrac{}{}{0pt}{}{0\leq s+t\leq\min(k,n_{+})}{k\leq s+2t}%
%EndExpansion
}(q^{3/2}\lambda_{+}^{1/2})^{m-k}\,q^{-2(m-k)n_{+}+2(k-s-t)(n_{3/0}%
-\,m+k)}\nonumber\\
&  \times\sum_{u=0}^{s}q^{-u+2u(n_{3/0}-\,m+2t+s)}\,(d_{q})_{s,t}^{k,n_{+}%
}\,[[m-k]]_{q^{2}}!\nonumber\\
&  \qquad\times%
%TCIMACRO{\QATOPD{[}{]}{n_{3/0}}{m-k}}%
%BeginExpansion
\genfrac{[}{]}{0pt}{}{n_{3/0}}{m-k}%
%EndExpansion
_{q^{2}}%
%TCIMACRO{\QATOPD{[}{]}{m}{k}}%
%BeginExpansion
\genfrac{[}{]}{0pt}{}{m}{k}%
%EndExpansion
_{q^{2}}\,(\hat{r}^{2})^{n_{r}}(X^{+})^{n_{+}-\,s-t+u}\nonumber\\
&  \qquad\times(X^{3/0})^{n_{3/0}-\,m+s+2t-u}\,S_{s,u}^{0}(\hat{r}^{2}%
,X^{3/0})\,(X^{-})^{n_{-}+\,m-s-t+u}.\nonumber
\end{align}
As usual, this expression enables us to read off the corresponding action on
commutative functions, i.e.
\begin{align}
&  (T^{-})^{m}\triangleright f(\hat{r}^{2},x^{+},x^{3/0},x^{-}) \label{T-fMin}%
\\
&  =\sum_{k=0}^{m}\sum_{s=0}^{k}\,\sum_{(k-s)/2\leq t\leq k-s}\,\sum_{u=0}%
^{s}q^{3m/2-k-t-u}\,(\lambda_{+}^{1/2})^{m-2t}\nonumber\\
&  \times q^{(k-s-t)(2(s+t)-k-m+1)-2u(m-2t-s)}\,%
%TCIMACRO{\QATOPD{[}{]}{m}{k}}%
%BeginExpansion
\genfrac{[}{]}{0pt}{}{m}{k}%
%EndExpansion
_{q^{2}}\nonumber\\
&  \times(x^{+})^{u}(x^{3/0})^{k+s+2t-u}\,S_{s,u}^{0}(r^{2},x^{3/0}%
)\,(x^{-})^{m-s-t+u}\nonumber\\
&  \times(\hat{T}_{q}^{+})_{s,t,k-s-t}\,(D_{q^{-2}}^{+})^{s+t}(D_{q^{2}}%
^{3/0})^{m-k}\,f(q^{-2(m-k)}x^{+},q^{2(k-s-t+u)}x^{3/0}),\nonumber
\end{align}
where, for brevity, we have introduced the operator%
\begin{align}
&  (\hat{T}_{q}^{+})_{s,l,i}\,f(r^{2},x^{+},x^{3/0},x^{-})\\
&  \equiv\sum_{%
%TCIMACRO{\QATOP{s_{1}+\ldots+s_{i+1}=s}{l_{1}+\ldots+l_{i+1}=l}}%
%BeginExpansion
\genfrac{}{}{0pt}{}{s_{1}+\ldots+s_{i+1}=s}{l_{1}+\ldots+l_{i+1}=l}%
%EndExpansion
}\Big (\prod_{k=1}^{i}q^{-2\sum_{j=1}^{k}(s_{j}+l_{j})}\,[[l-i+k-l_{1}%
-\ldots-l_{k}]]_{q^{2}}\Big )\nonumber\\
&  \times\sum_{%
%TCIMACRO{\QATOP{0\leq t_{\alpha}\leq l_{\alpha}}{\alpha=1,\ldots,i+1}}%
%BeginExpansion
\genfrac{}{}{0pt}{}{0\leq t_{\alpha}\leq l_{\alpha}}{\alpha=1,\ldots,i+1}%
%EndExpansion
}\Big (\prod_{k=1}^{i+1}q^{2t_{k}\sum_{u=1}^{k-1}(s_{u}+t_{u})}\,(P_{q}%
)_{t_{k}}^{s_{k},l_{k}}\Big )f(q^{2\sum_{u=1}^{i+1}t_{u}}x^{+}).\nonumber
\end{align}

Following the same line of arguments, we can also derive explicit formulae for
the action of powers of some more Lorentz generators. For the computation of
braided products in Sec. \ref{MinSpa} we need to know
\begin{align}
&  (S^{1})^{m}\triangleright f(r^{2},x^{+},x^{3/0},x^{-})\\
&  =(-q^{1/2}\lambda_{+}^{-1/2})^{m}\,(x^{3/0})^{m}\,(D_{q^{2}}^{+}%
)^{m}\,f(q^{-m}x^{+},q^{-m}x^{3/0},q^{m}x^{-}),\nonumber\\[0.16in]
&  (T^{2})^{m}\triangleright f(r^{2},x^{+},x^{3/0},x^{-})\\
&  =(q\lambda_{+})^{-m/2}\,(x^{3/0})^{m}\,(D_{q^{2}}^{-})^{m}\,f(q^{m}%
x^{+},q^{-m}x^{3/0},q^{-m}x^{-}),\nonumber\\[0.16in]
&  (\tau^{1})^{m}\triangleright f(r^{2},x^{+},x^{3/0},x^{-})\label{WirkTau}\\
&  =\sum_{k=0}^{\infty}(-\lambda_{+}^{-1}\lambda^{2})^{k}\,q^{-k(k-1)}\,%
%TCIMACRO{\QATOPD{[}{]}{m}{k}}%
%BeginExpansion
\genfrac{[}{]}{0pt}{}{m}{k}%
%EndExpansion
_{q^{2}}\,(x^{3/0})^{2k}\nonumber\\
&  \times(D_{q^{2}}^{+})^{k}(D_{q^{2}}^{-})^{k}\,f(q^{m}x^{+},x^{m-2k}%
x^{3/0},q^{-m}x^{-}),\nonumber\\[0.16in]
&  (\sigma^{2})^{m}\triangleright f(r^{2},x^{+},x^{3/0},x^{-})=f(q^{-m}%
x^{+},q^{-m}x^{3/0},q^{m}x^{-}),\label{WirkSigma}\\
&  (\tau^{3})^{m}\triangleright f(r^{2},x^{+},x^{3/0},x^{-})=f(q^{-4m}%
x^{+},q^{4m}x^{-}).\nonumber
\end{align}

\end{document}